\documentclass[10pt,journal,compsoc]{IEEEtran}
\ifCLASSOPTIONcaptionsoff
    \usepackage[nomarkers]{endfloat}
\let\MYoriglatexcaption\caption
\renewcommand{\caption}[2][\relax]{\MYoriglatexcaption[#2]{#2}}
\fi
\usepackage[T1]{fontenc}
\usepackage{url}
\usepackage{subfigure}
\usepackage{color}
\usepackage{dsfont}
\usepackage{bm}
\usepackage[pdftex]{graphicx}
\usepackage{array,amsmath,amssymb,amsfonts}
\usepackage{amsthm}
\usepackage{subeqnarray}
\usepackage{cases}
\usepackage{booktabs}
\usepackage{multirow}
\usepackage[table,xcdraw]{xcolor}
\usepackage[linesnumbered,ruled,vlined]{algorithm2e}
\makeatletter
\newcommand{\removelatexerror}{\let\@latex@error\@gobble}
\makeatother

\SetCommentSty{mycommfont}

\renewcommand{\vec}[1]{\boldsymbol{#1}}
\DeclareMathOperator*{\argmin}{argmin}
\DeclareMathOperator*{\argmax}{argmax}
\DeclareMathOperator*{\diag}{diag}
\newtheorem{theorem}{\textbf{Theorem}}

\newtheorem{proposition}{\textbf{Proposition}}
\newtheorem{definition}{\textbf{Definition}}
\def\BibTeX{{\rm B\kern-.05em{\sc i\kern-.025em b}\kern-.08em
    T\kern-.1667em\lower.7ex\hbox{E}\kern-.125emX}}
\IEEEoverridecommandlockouts
\usepackage{siunitx} % Required for alignment

\begin{document}
\title{Scheduling Multi-Server Jobs with Sublinear Regrets via Online Learning
% \thanks{\color{black}Identify applicable funding agency here.}
}

\author{
        Hailiang~Zhao,
        Shuiguang~Deng,~\IEEEmembership{Senior~Member,~IEEE,} 
        Zhengzhe~Xiang,
        Xueqiang~Yan,
        Jianwei~Yin,
        Schahram~Dustdar,~\IEEEmembership{Fellow,~IEEE},
        and~Albert~Y.~Zomaya,~\IEEEmembership{Fellow,~IEEE}%
\IEEEcompsocitemizethanks{
  \IEEEcompsocthanksitem H. Zhao and S. Deng are with Hainan Institute of Zhejiang University (Sanya 572025, China) and the College of Computer Science and Technology, Zhejiang University (Hangzhou 310027, China). E-mail: \{hliangzhao, dengsg\}@zju.edu.cn
  \IEEEcompsocthanksitem Z. Xiang is with the School of Computer Science and Technology, Hangzhou City University, Hangzhou 310015, China. E-mail: xiangzz@zucc.edu.cn
  \IEEEcompsocthanksitem X. Yan is with Huawei Technologies Co Ltd, Shanghai 201206, China. E-mail: yanxueqiang1@huawei.com
  \IEEEcompsocthanksitem J. Yin is with the College of Computer Science and Technology, Zhejiang University, Hangzhou 310027, China. E-mail: zjuyjw@zju.edu.cn
  \IEEEcompsocthanksitem S. Dustdar is with the Distributed Systems Group, Technische Universität Wien, 1040 Vienna, Austria. E-mail: dustdar@dsg.tuwien.ac.at
  \IEEEcompsocthanksitem A. Y. Zomaya is with the School of Computer Science, University of Sydney, Sydney, NSW 2006, Australia. E-mail: albert.zomaya@sydney.edu.au
  \IEEEcompsocthanksitem Shuiguang Deng is the corresponding author.}%
  % \thanks{Manuscript received November --, 2020; revised November --, 2020.}
}

\IEEEtitleabstractindextext{%
\begin{abstract}
    Multi-server jobs that request multiple computing resources and hold onto them during their execution dominate modern computing clusters. {\color{black}When allocating the multi-type resources to several co-located multi-server jobs simultaneously in online settings, it is difficult to make the tradeoff between the parallel computation gain and the internal communication overhead, apart from the resource contention between jobs. To study the computation-communication tradeoff, we model the computation gain as the speedup on the job completion time when it is executed in parallelism on multiple computing instances, and fit it with utilities of different concavities. Meanwhile, we take the dominant communication overhead as the penalty to be subtracted. To achieve a better gain-overhead tradeoff, we formulate an cumulative reward maximization program and design an online algorithm, named \textsc{OgaSched}, to schedule multi-server jobs. \textsc{OgaSched} allocates the multi-type resources to each arrived job in the ascending direction of the reward gradients. It has several parallel sub-procedures to accelerate its computation, which greatly reduces the complexity. We proved that it has a sublinear regret with general concave rewards.} We also conduct extensive trace-driven simulations to validate the performance of \textsc{OgaSched}. The results demonstrate that \textsc{OgaSched} outperforms widely used heuristics by $11.33\%$, $7.75\%$, $13.89\%$, and $13.44\%$, respectively.
\end{abstract}

\begin{IEEEkeywords}
    Multi-server job, online gradient ascent, online scheduling, regret analysis.
\end{IEEEkeywords}}

\maketitle

\section{Introduction}\label{s1}
In today's computing clusters, whether in the cloud data centers or at the network edge, many jobs request multiple resources (CPUs, GPUs, etc.) simultaneously and hold onto them during their executions. For example, graph computations \cite{10.1145/2806777.2806849}, federated learning \cite{8885054}, distributed DNN model trainings \cite{DL2}, etc. In this paper, we refer to these jobs as \textit{multi-server jobs} \cite{zero-queue,9925644}. Multi-server jobs of diverse resource requirements arrive at the cluster online, which puts great pressure to current resource allocation policies to achieve a high computation efficiency.

When allocating the multi-type resources to several co-located multi-server jobs simultaneously in online settings, it is difficult to make the tradeoff between \textit{the parallel computation gain} and \textit{the internal communication overhead}, apart from the resource contention between jobs. Here the parallel computation gain refers to the speedup on the job completion time when it is executed in parallelism on multiple computing instances, which could be modeled with a function of the allocated multi-type resources \cite{bao2018online}. Correspondingly, the internal communication overhead refers to the cost caused by non-computation operations such as data synchronization, averaging, message passing, etc., between the distributed workers. To achieve better computation efficiency, we need to consider the following key challenges.

\begin{itemize}
    \item {\color{black}\textit{Resource contention with service locality.} With service locality, a multi-server job can only be processed by a subset of computing instances where the resource requirements, session affinity \cite{carrion2022kubernetes}, and other obligatory constraints are satisfied. When several multi-server jobs arrive simultaneously, how to allocate the limited resources to them without degenerating the computation efficiency is challenging. }
    \item {\color{black}\textit{Unknown arrival patterns of jobs.} In real-life scenarios, the resource allocation should be made online without the knowledge of future job arrivals. The lack of information on the problem space could lead to a solution far from the global optimum.}
    \item {\color{black}\textit{The parallel computation gain does not increase in a linear rate with the quantity of allocated resources.} For instance, in distributed DNN model training or federated learning, adding workers (that request more resources) does not improve the training speed linearly \cite{DL2,bao2018online}. This is because the overhead of all-reduce operation between workers or the averaging of local gradients increase with the number of participated workers, especially when the workers are distributed in different machines and communicate with each other through network \cite{258953,280874,280776}. Compared with high-speed intra-node communication channels such as NVLink, the inter-node bandwidth through NIC is relatively much slower. Another example is graph computation. Without a well-designed graph partition policy, the speedup of message-passing between graph nodes can be significantly slowed down \cite{10.1145/2806777.2806849,180822}.}
    \item \textit{The type of resource which dominates the communication overhead varies to different job types.} {\color{black}For example, in graph computation jobs, the dominant communication overhead lies in the internal input-output data transferring between the interdependent CPU- and memory-intensive tasks \cite{decima}. However, the dominant overhead of the distributed training of DNNs lies in the data averaging and synchronizing between the GPU-intensive workers through network \cite{BSP}. This variety greatly complicates the theoretical analysis for the gain-overhead tradeoff. }
\end{itemize}

Despite the vast literature on the online resource allocation algorithms \cite{DL2,bao2018online,BSP,8737465,8737612,8737370,8486340,yu2021sum}, {\color{black}their model formulation and theoretical analysis which places emphasis on the gain-overhead tradeoff is limited. To fill the theoretical gap, in this paper, we propose an online scheduling algorithm, termed as \textsc{OgaSched}, to \textit{learn} to allocate multi-type resources to co-located multi-server jobs online to maximize the overall computation efficiency.} We try to analyze the tradeoff in a generic way. The generality is embodied in the following points. First of all, different from the specific works on deep learning jobs \cite{bao2018online,BSP,DL2} or query jobs \cite{decima}, we allow different types of multi-server jobs to \textit{co-locate} in the cluster which consists of heterogeneous computing resources. Different job types can have different resource requirements while different computing instances can be equipped with diverse quantities and types of  resources. Secondly, we adopt general zero-startup non-decreasing utility functions to model the parallel computation gain in terms of the job completion time. {\color{black}Compared to existing literature, we allow the utilities to be diverse in their \textit{level of concavity}.} Specifically, we provide both analysis and experiments on linear, polynomial, logarithmic, and reciprocal utilities. Thirdly, we makes no assumptions on the arrival patterns of multi-server jobs. \textsc{OgaSched} requests no knowledge on the job arrival distributions but tries to learn them to make better scheduling decisions. 

In our model formulation, the computation efficiency is modeled in the way of cumulative reward. Time is slotted, and the cumulative reward is obtained by summing up the reward in each time slot, where a single-time reward is a linear aggregation of each job's reward. Further, a job's reward at each time is designed as the achieved parallel computation gain aggregated over the allocated resources minus the penalty introduced by the dominant communication overhead. At each time, \textsc{OgaSched} allocates resources to each arrived job in the direction that \textit{makes the gradient of the reward increase}. \textsc{OgaSched} is capable of handling high dimensional inputs {\color{black}in stochastic scenarios with unpredictable behaviors}. We adopt regret, i.e., the gap on the cumulative reward between the proposed online algorithm and the offline optimum achieved by an oracle \cite{anderson2008theory}, to analyze the performance lower bound of \textsc{OgaSched}. We prove that, \textsc{OgaSched} has a State-of-the-Art (SOTA) regret, which is sublinear with the time slot length and the number of job types. This work fulfills one of the key deficiencies of the past works in the modeling and analysis of the gain-overhead tradeoff for multi-server jobs. The contributions are summarized as follows. 

\begin{itemize}
    \item {\color{black}We systematically study the resource allocation of co-located multi-server jobs in terms of the tradeoff between the parallel computation gains and the internal communication overheads. Our study is general in scenario settings and it sufficiently takes the characters of the diminishing marginal effect of gains into consideration. }
    
    \item {\color{black}We propose an algorithm, i.e., \textsc{OgaSched}, to learn to strike a balanced computation-communication tradeoff. \textsc{OgaSched} has no assumptions on the job arrival patterns. With a nice setup (defined in Sec. \ref{s3.1}), \textsc{OgaSched} achieves a SOTA regret $\mathcal{O} \big( \mathcal{H}_{\mathcal{G}} \cdot \sqrt{T} \big)$ for general concave non-linear rewards, where $T$ is the time slot length, and $\mathcal{H}_\mathcal{G}$ (formally defined in \eqref{hg}) is parameter that characterizes the bipartite graph model.} \textsc{OgaSched} is accelerated by well-designed parallel sub-procedures. The parallelism helps yield a complexity of $\mathcal{O} \big( \log(K) \big)$, where $K$ is the number of resource types. 
    
    \item We conduct extensive trace-driven simulations to validate the performance of \textsc{OgaSched}. The simulation results show that \textsc{OgaSched} outperforms widely used heuristics including \textsc{DRF} \cite{DRF}, \textsc{Fairness}, \textsc{BinPacking}, and \textsc{Spreading} by $11.33\%$, $7.75\%$, $13.89\%$, and $13.44\%$, respectively. We also provide large-scale validations.
\end{itemize}

The rest of this paper is organized as follows. We formulate the online scheduling problem for multi-server jobs in Sec. \ref{s2}. We then present the design details of \textsc{OgaSched} with regret analysis and discuss its extensions in Sec. \ref{s3}. We demonstrate the experimental results in Sec. \ref{s4}, and discuss related works in Sec. \ref{s5}. Finally, we conclude this paper in Sec. \ref{s6}.
% ======================================================================================================================================================

\section{Bipartite Scheduling with Regrets}\label{s2}
We consider a cluster of heterogenous computing instances serving several types of multi-server jobs. Here the computing instances can be VMs in clouds, or local servers at the network edge. The computing instances work collaboratively to provide resources to serve the considered jobs. Different computing instances are equipped with different types and quantities/specifications of resources, including CPU cores, memory, bandwidth, GPUs, etc. Jobs of different types can have different demands on them. 
Key notations used in this paper are summarized in Tab. \ref{tab1}.

\begin{table}[htbp]   
    \begin{center}
        \caption{\label{tab1}Summary of key notations.}   
        \begin{tabular}{l|l}    
            \toprule
            {\textsc{Notation}}& {\textsc{Description}}\\[+0.1mm]
            \midrule
            $\mathcal{T}$ & Time horizon of length $T$\\[+0.7mm]
            $\mathcal{G} = (\mathcal{L}, \mathcal{R}, \mathcal{E})$ & The bipartite graph\\[+0.7mm]
            $l \in \mathcal{L}$ & A job type (port)\\[+0.7mm]
            $r \in \mathcal{R}$ & A computing instance\\[+0.7mm]
            $(l,r) \in \mathcal{E}$ & The edge (channel) between $l$ and $r$\\[+0.7mm]
            $\forall r: \mathcal{L}_r$ & The set of job types connect to $r$\\[+0.7mm]
            $\forall l: \mathcal{R}_l$ & The set of computing instances connect to $l$\\[+0.7mm]
            $\vec{x}(t)$ & The job arrival status at time $t$\\[+0.7mm]
            $\vec{y}(t)$ & The scheduling decision at time $t$\\[+0.7mm]
            $\mathcal{K}$ & The set of different types of resources\\[+0.7mm]
            $\forall l: \vec{a}_l$ & Resource requirements of type-$l$ job\\[+0.7mm]
            $\forall r, k: c_r^k$ & The number of type-$k$ resources equipped by $r$\\[+0.7mm]
            $q\big( \vec{x}(t), \vec{y}(t) \big)$ & The reward of time $t$\\[+0.7mm]
            $\forall k: f_k(\cdot)$ & Computation gain of type-$k$ devices\\[+0.7mm]
            $\forall k: \beta_k \in [0,1]$ & Coefficient of type-$k$ communication overhead\\[+0.7mm]
            \bottomrule   
        \end{tabular}  
    \end{center}
\end{table}

\subsection{Online Bipartite Scheduling}\label{s2.1}
\begin{figure}[htbp]
    \centerline{\includegraphics[width=3in]{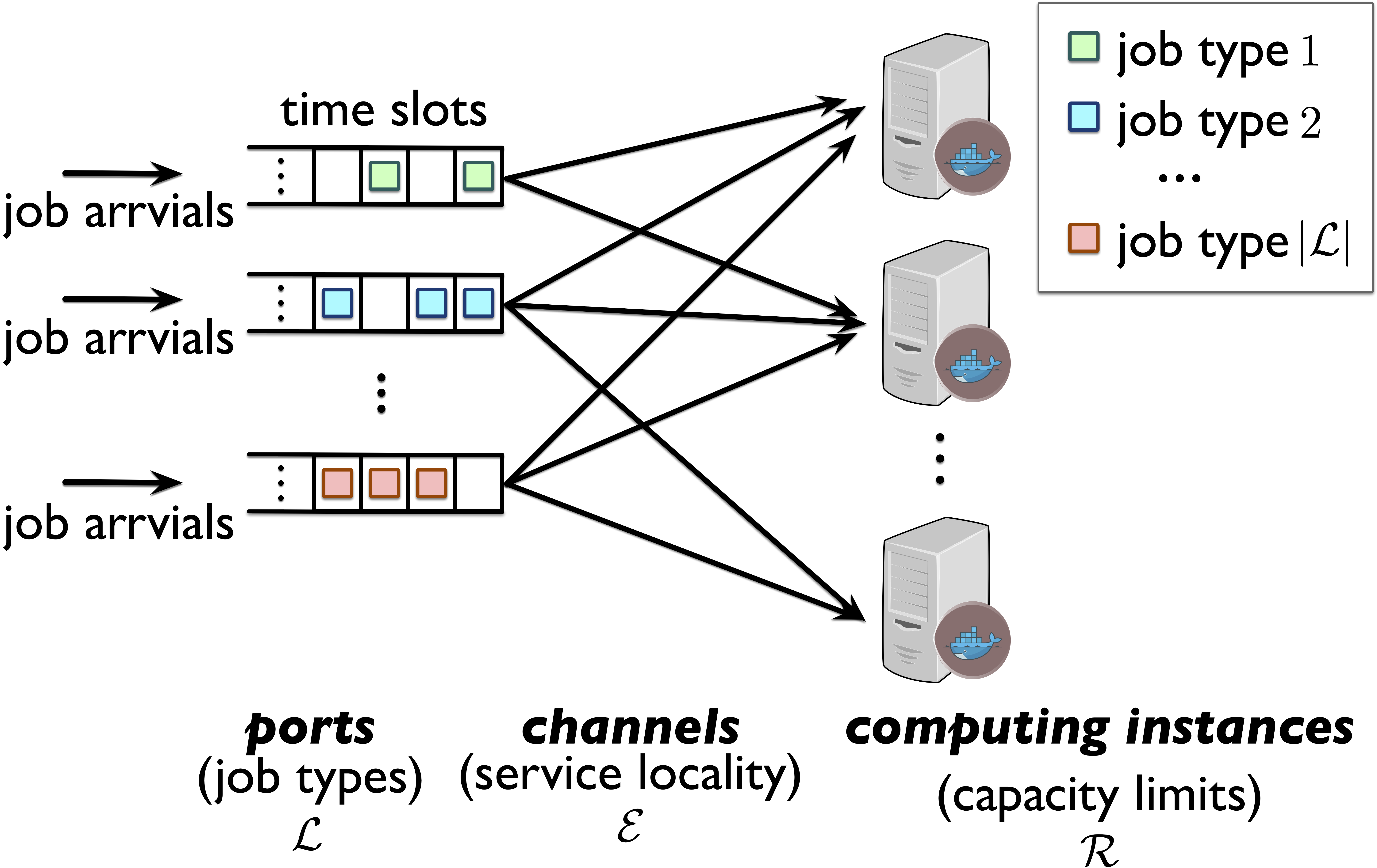}}
    \caption{The bipartite graph model for online job scheduling.}
    \label{fig1}
\end{figure}

We use a bipartite graph $\mathcal{G} = (\mathcal{L}, \mathcal{R}, \mathcal{E})$ to model the job-server constraints, 
as shown in Fig. \ref{fig1}. In graph $\mathcal{G}$, $\mathcal{L}$ is the set of job types and indexed by $l$ 
while $\mathcal{R}$ is the set of computing instances and indexed by $r$. The connections between the job types and the 
computing instances are recorded in $\mathcal{E}$. Because of the job-server constraints, type-$l$ job may only be 
served by a subset of $\mathcal{R}$. We denote the subset by
\begin{flalign}
    \mathcal{R}_l = \Big\{ r \in \mathcal{R} \mid (l,r) \in \mathcal{E} \Big\}.
\end{flalign}
Similarly, we use
\begin{flalign}
    \mathcal{L}_r = \Big\{ l \in \mathcal{L} \mid (l,r) \in \mathcal{E} \Big\}.
\end{flalign}
to represent the set of job types that connect 
to computing instance $r$. We designate each job type $l \in \mathcal{L}$ as \textit{port} and each connection $(l,r) \in \mathcal{E}$ 
as \textit{channel}. $\mathcal{G}$ is called right $d$-regular \textit{iff} 
the indegree of each right vertex is $d$, i.e., $\forall r \in \mathcal{R}, |\mathcal{L}_r| = d$.

Time is discretized, and at each time $t \in \mathcal{T} \triangleq \{1, ..., T\}$, from each port, at most one job yields. 
Let us denote by
\begin{equation}
    \vec{x}(t) = \Big[x_l(t)\Big]_{l \in \mathcal{L}} \in \big\{0, 1\big\}^{|\mathcal{L}|}
\end{equation}
the job arrival status at time $t$. We do not make any assumption on the job arrival patterns or distributions.
The cluster has $K$ types of resources, and computing instance $r$ has 
$c_r^k$ type-$k$ resources, where $k \in \mathcal{K} \triangleq \{1, 2, .., K\}$. For each type-$l$ job, we denote its maximum requests 
on each resource by $\vec{a}_l = [a_l^k]_{k \in \mathcal{K}} \in \mathbb{N}^{|\mathcal{K}|}$. At time $t$, we use 
\begin{equation}
    \vec{y}(t) = \Big[y_{(l,r)}^k(t) \Big]_{l \in \mathcal{L}, r \in \mathcal{R}_l, k \in \mathcal{K}} \in 
    \mathbb{R}^{\sum_{l \in \mathcal{L}} |\mathcal{R}_l| \times K}_{\geq 0}
\end{equation}
to denote the scheduling decision. Here we allow $y_{(l,r)}^k(t)$ to be fractional. Taking GPU as example, Machine-Learning-as-a-Service (MLaaS) platforms support GPU sharing in a space- and time-multiplexed manner by intercepting CUDA APIs \cite{time-sharing1,time-sharing2,276938}.

The first constraint is that, through each channel, a job should not be allocated with resources more than it requires. Formally, we have 
\begin{equation}
    0 \leq y_{(l,r)}^k(t) \leq a_l^k, \forall l,r,k,t.
    \label{cons1}
\end{equation}
The second constraint $\vec{y}(t)$ should satisfy is that, the resources allocated out from any computing instance $r$
should not more than it has:
\begin{equation}
    \sum_{l \in \mathcal{L}_r} y_{(l,r)}^k(t) \leq c_r^k, \forall r, k, t.
    \label{cons2}
\end{equation}
We denote by $\mathcal{Y} \triangleq \big\{ \vec{y} \in \mathbb{R}^{\sum_{l \in \mathcal{L}} |\mathcal{R}_l| \times K} 
\mid \eqref{cons1} \textrm{ and } \eqref{cons2} \textrm{ hold} \big\}$ to represent the solution space from here on. 

\subsection{Computation-Communication Tradeoff}\label{s2.2}
The performance metric we use for online bipartite scheduling is designed as the gain obtained by the parallel computation through multi-type resources minus the penalty 
introduced by the dominant communication overheads. Specifically, we denote by $q_l\big(\vec{x}(t), \vec{y}(t)\big)$ the 
reward of port $l$ at time $t$ , and it is formulated as 

\begin{flalign}
    q_l\big(\vec{x}(t), \vec{y}(t)\big) = x_l(t) 
    \bigg[ &\sum_{k \in \mathcal{K}} f_k \Big(\sum_{r \in \mathcal{R}_l} y_{(l,r)}^k(t)\Big) - \nonumber \\
    &\max_{k \in \mathcal{K}} \Big\{ \beta_k \sum_{r \in \mathcal{R}_l} y_{(l,r)}^k(t) \Big\} \bigg].
    \label{reward}
\end{flalign}
In this formulation, the first part, $\sum_{k \in \mathcal{K}} f_k (\sum_{r \in \mathcal{R}_l} y_{(l,r)}^k(t) )$, is the parallel computation gain, which is linearly aggregated over each type of resource, in proportional to each resource's weight. {\color{black}Jobs of different types can have different combinations of weights.} $f_k (\cdot)$ is the gain achieved by $\sum_{r \in \mathcal{R}_l} y_{(l,r)}^k(t)$ type-$k$ resources collaboratively, where $f_k (\cdot)$ is a zero-startup concave utility defined in $\mathbb{R}_{\geq 0}$. Note that $\sum_{r \in \mathcal{R}_l} y_{(l,r)}^k(t)$ is the quota of the type-$k$ resources allocated to the type-$l$ job at $t$. {\color{black}As we have analyzed before, $\{f_k(\cdot)\}_{k \in \mathcal{K}}$ are non-decreasing concave functions because the marginal effect of parallel computation decreases successively when increasing participated resources \cite{marginal-effect,9328612}.} We expect $\{f_k(\cdot)\}_{k \in \mathcal{K}}$ to be \textit{continuously differentiable} because it helps design a policy that yields a nice lower bound of the reward. The details will be demonstrated in Sec. \ref{s3.3}. If $\{f_k(\cdot)\}_{k \in \mathcal{K}}$ are not differentiable everywhere, we can apply subgradient ascent-related techniques in the policy design. The second part in \eqref{reward} is $\max_{k \in \mathcal{K}} \big\{ \beta_k \sum_{r \in \mathcal{R}_l} y_{(l,r)}^k(t) \big\}$, which reflects the dominant weighted communication overheads over different types of resources. {\color{black}For example, in federated learning at the edge, the dominant communication overhead lies in the averaging and synchronizing of data between each edge server over the network \cite{bao2022federated}. Another example is graph computation, in which the job is organized into a direct acyclic graph (DAG), and the dominant communication overhead falls into the data \& message passing between CPU- and memory-intensive tasks \cite{decima}.} $\{\beta_k\}_{k \in \mathcal{K}}$ are the coefficients to balance the gain and the overhead. W.L.O.G., we set each $\beta_k \in [0,1]$. Theoretically, the second part of \eqref{reward} is a penalty, the minimization of which guides the scheduling decisions to balance the communication overheads of different device types. Our reward design encourages each job to be served with the balance between the computation gain and the communication overhead being achieved. 

\subsection{Regret Minimizing}
Based on the above, we define the overall reward at time $t$ as the linear aggregation over each port: 
\begin{flalign}
    q \big(\vec{x}(t), \vec{y}(t)\big) = \sum_{l \in \mathcal{L}} q_l\big(\vec{x}(t), \vec{y}(t)\big).
    \label{q}
\end{flalign}
The cumulative reward of scheduling policy $\pi$ over the time horizon $\mathcal{T}$ is obtained by summing up 
the rewards obtained at each time until $T$: 
\begin{flalign}
    Q^\pi \Big( \{\vec{x}(t) \}_1^T, \{\vec{y}(t)\}_1^T \Big) = \sum_{t \in \mathcal{T}} q \big(\vec{x}(t), \vec{y}(t)\big),
\end{flalign} 
where the scheduling decisions $\{\vec{y}(t)\}_1^T$ are made under the guidance of policy $\pi$. In the following, we just use 
$Q$ and drop the superscript $\pi$ for simplification.

We do not make any assumption on the distribution of the job arrival trajectory $\{\vec{x}(t)\}_1^T$.
To obtain a non-trivial performance measure, we cast the multi-server bipartite scheduling problem into the framework of online learning, which prompts us to compare the performance of the online policy $\pi$ with the best offline stationary policy $\pi^*$ \cite{online-learning-1,fundamental}. Let us denote by $\vec{y}^*$ the optimal offline stationary resource allocation decision guided by policy $\pi^*$, i.e., 
\begin{flalign}
    \vec{y}^* = \arg \sup_{\vec{y} \in \mathcal{Y}} Q \Big( \{\vec{x}(t)\}_1^T, \vec{y} \Big),
\end{flalign}
Physically, $\vec{y}^*$ is the optimal stationary resource reservation decisions for each port whatever the actual job arrival status $\vec{x}(t)$ is. Formally, we define the regret $R_T^{\pi}\big(\{ \vec{x}(t) \}_1^T\big)$ for the job arrival trajectory $\{\vec{x}(t)\}_1^T$ as
\begin{equation*}
    R_T^{\pi}\Big(\{ \vec{x}(t) \}_1^T\Big) \triangleq 
    Q \Big( \big\{\vec{x}(t)\big\}_1^T, \vec{y}^* \Big) - Q \Big( \big\{\vec{x}(t)\big\}_1^T, \big\{\vec{y}(t)\big\}_1^T \Big).
\end{equation*}
The regret of policy $\pi$ is further defined as the maximum regret achieved over every possible job arrival trajectory:
\begin{equation}
    R_T^\pi \triangleq \sup_{ \forall \{\vec{x}(t)\}_1^T } R_T^{\pi}\Big( \big\{ \vec{x}(t) \big\}_1^T \Big).
\end{equation}
Our goal is to find a policy $\pi$, under which a sequence of bipartite scheduling decisions $\{\vec{y}(t)\}_1^T$ is yielded, 
to minimize $R_T^\pi$.
% ======================================================================================================================================================

\section{Online Gradient Ascent}\label{s3}
To minimize the regret $R_T^\pi$, we resort to an online variant of the gradient-based methods, online gradient ascent (OGA) 
\cite{oga-base}. A series of recent works have demonstrated that OGA achieves the best possible regret for online caching problems in 
different network settings when the rewards are linear \cite{fundamental,23,caching-follow-1,caching-follow-2}. 
In this paper, we extend OGA to the online bipartite scheduling problem for multi-server jobs with non-linear rewards. Before 
presenting the design details, we first give some preliminary definitions and analysis.

 \subsection{Preliminaries}\label{s3.1}
\begin{definition}
    \textsc{Nice Setup}. If all the utilities $\{ f_k \}_{k \in \mathcal{K}}$ are $(i)$ linearly 
    separable over computing instances, i.e., 
    \begin{equation}
        f_k \Big(\sum_{r \in \mathcal{R}_l} y_{(l,r)}^k\Big) = \sum_{r \in \mathcal{R}_l} f_r^k \Big( y_{(l,r)}^k \Big),
    \end{equation}
    and each concave utility $f_r^k(\cdot)$ is $(ii)$ continuously differentiable in 
    $\mathbb{R}_+$, and $(iii)$ there exist $\varpi_r^k > 0$ such that
    \begin{equation}
        (f_r^k)' (0) \leq \varpi_r^k, \forall r,k,
    \end{equation}
    we say this is a nice setup.
\end{definition}
% With a nice setup, each separated utility $f_r^k (\cdot)$ inherits the mathematical properties of $f_k (\cdot)$. 
The following proposition demonstrates the property of the regret minimization problem, which will be used in the design and analysis of \textsc{OgaSched}.
\begin{proposition}
    \textsc{Convexity}. $(i)$ The feasible solution space $\mathcal{Y}$ is convex. 
    $(ii)$ With a nice setup, at each time $t$, the single-slot reward function $q\big( \vec{x}(t), \vec{y}(t) \big)$ is a 
    concave function of $\vec{y}(t)$.
\end{proposition}
\begin{proof}
    In the following proof, we just drop $(t)$ from $\vec{x}(t)$ and $\vec{y}(t)$ for simplification. Besides, we only prove 
    the case that $\mathcal{G}$ is right $d$-regular and $d = |\mathcal{L}|$. The left cases can be easily proved with the 
    same techniques used in this proof.

    We firstly prove $(i)$. To do this, let us arrange the vector $\vec{y}$ as 
    \begin{equation}
        \vec{y} = \Big[ \underbrace{\vec{y}^1}_{k=1}; \underbrace{\vec{y}^2}_{k=2}; ...; \underbrace{\vec{y}^K}_{k=K} \Big],
    \end{equation}
    where $\vec{y}^k \in \mathbb{R}^{(|\mathcal{L}| \times |\mathcal{R}|)}$ is arranged as
    \begin{equation}
        \vec{y}^k = \Big[ 
            \underbrace{y_{(1, 1)}^k; ...; y_{(1, |\mathcal{R}|)}^k}_{l=1}; ...; 
            \underbrace{y_{(|\mathcal{L}|, 1)}^k; ...; y_{(|\mathcal{L}|, |\mathcal{R}|)}^k}_{l=|\mathcal{L}|} \Big].
    \end{equation}
    With this arrangement, the vector representation of \eqref{cons1} is
    \begin{equation}
        \vec{0} \leq \vec{y} \leq \vec{a},
    \end{equation}
    where $\vec{a} = \big[ \vec{a}^1; ...; \vec{a}^K \big]$, and 
    \begin{equation}
        \vec{a}^k = \Big[ 
            \underbrace{a_1^k; ...; a_1^k}_{\textrm{of size }|\mathcal{R}|}; ...; 
            \underbrace{a_{|\mathcal{L}|}^k; ...; a_{|\mathcal{L}|}^k}_{\textrm{of size }|\mathcal{R}|} \Big].
    \end{equation}
    Similarly, we want to construct a matrix $\mathbf{B}$ and a vector $\vec{c}$ for the vector representation of \eqref{cons2}. 
    $\forall k \in \mathcal{K}$, we design $\mathbf{B}' \in \mathbb{R}^{|\mathcal{R}| \times (|\mathcal{L}| \times |\mathcal{R}|)}$ as
    \begin{equation}
        \big[ \mathbf{B}' \big]_{ij} = 
        \left\{
            \begin{array}{ll}
                1 & (j - i) \mid |\mathcal{R}| \\
                0 & \textrm{o.w.}
            \end{array}
        \right.
    \end{equation}
    and $\vec{c}^k = \big[ c_1^k; ...; c_{|\mathcal{R}|}^k \big]$. Then, we have
    \begin{equation}
        \mathbf{B}' \vec{y}^k \leq \vec{c}^k, \forall k.
        \label{before-transform}
    \end{equation}
    We can transform \eqref{before-transform} into 
    \begin{equation}
        \mathbf{B}' \vec{y}^k + \sum_{k' \neq k} \mathbf{O} \vec{y}^{k'} \leq \vec{c}^k, \forall k,
    \end{equation}
    where $\mathbf{O}$ is a zero matrix. As a result, \eqref{cons2} is equivalent to 
    \begin{equation}
        \mathbf{B} \vec{y} \leq \vec{c},
    \end{equation}
    where $\vec{c} = \big[ \vec{c}^1;...;\vec{c}^K \big] \in \mathbb{R}^{(|\mathcal{R}| \times K)}$, and 
    \begin{equation*}
        \mathbf{B} = \diag \big( \mathbf{B}' \big) \in \mathbb{R}^{(|\mathcal{R}| \times K) \times (|\mathcal{L}| \times |\mathcal{R}| \times K)}.
    \end{equation*}
    The above analysis leads to $\mathcal{Y} = \big \{\vec{y} \mid \vec{0} \leq \vec{y} \leq \vec{a}, \mathbf{B} \vec{y} \leq \vec{c} \big\}$ 
    being a polyhedron, which is well known to be a convex set.

    We now prove $(ii)$. Similarly, we try to find the vectorized representation of $q \big(\vec{x}, \vec{y}\big)$. To do this, 
    we define the operator $\vec{f}: \mathbb{R}^{|\mathcal{L}| \times |\mathcal{R}| \times K} \to \mathbb{R}^{|\mathcal{L}| \times |\mathcal{R}| \times K}$ as 
    \begin{equation}
        \vec{f} = \big[ \vec{f}^1; ...; \vec{f}^K \big],
    \end{equation}
    where 
    \begin{equation}
        \vec{f}^k = \Big[ 
            \underbrace{f_1^k; ...; f_{|\mathcal{R}|}^k}_{\textrm{make }|\mathcal{L}| \textrm{ replicas}}; ...; 
            \underbrace{f_1^k; ...; f_{|\mathcal{R}|}^k}_{l=|\mathcal{L}|} \Big].
    \end{equation}
    Then, the first part of \eqref{reward} can be transformed into 
    \begin{flalign}
        \sum_{l \in \mathcal{L}} x_l \sum_{k \in \mathcal{K}} f_k \Big(\sum_{r \in \mathcal{R}_l} y_{(l,r)}^k\Big) &=
        \sum_{l \in \mathcal{L}} \sum_{k \in \mathcal{K}} \sum_{r \in \mathcal{R}_l} x_l f_r^k \Big( y_{(l,r)}^k \Big) \nonumber \\
        &= \vec{\chi} \cdot \vec{f} (\vec{y}),
    \end{flalign}
    where 
    \begin{equation}
        \vec{\chi} = \Big[ 
            \underbrace{\underbrace{x_1; ...; x_1}_{l=1, \forall r}; ...; 
            \underbrace{x_{|\mathcal{L}|}; ...; x_{|\mathcal{L}|}}_{l=|\mathcal{L}|, \forall r}}_{k=1, \textrm{ make } K \textrm{ replicas}}; 
            ...; 
            \underbrace{x_1; ...; x_{|\mathcal{L}|}}_{k = K} \Big].
    \end{equation}
    For the second part of \eqref{reward}, we have
    \begin{equation}
        \sum_{l \in \mathcal{L}} x_l \max_{k \in \mathcal{K}} \Big\{ \beta_k \sum_{r \in \mathcal{R}_l} y_{(l,r)}^k \Big\} = 
        \sum_{l \in \mathcal{L}} x_l \beta_{k^*} \sum_{r \in \mathcal{R}_l} y_{(l,r)}^{k^*},
    \end{equation}
    where 
    \begin{equation}
        k^* = \argmax_{k \in \mathcal{K}} \Big\{ \beta_k \sum_{r \in \mathcal{R}_l} y_{(l,r)}^k \Big\}.
        \label{k-star}
    \end{equation}
    Without loss of generality, we assume that $k^* = 1$. Then, the second part can be represented as $\vec{\beta} \cdot \vec{y}$, where
    \begin{equation}
        \vec{\beta} = \Big[ 
            \underbrace{\underbrace{x_1 \beta_1; ...; x_1 \beta_1}_{l=1, \forall r}; ...; 
            \underbrace{x_{|\mathcal{L}|} \beta_1; ....; x_{|\mathcal{L}|} \beta_1}_{l=|\mathcal{L}|, \forall r}}_{k^* = 1}; 
            \underbrace{\vec{0}}_{\forall k \neq k^*} \Big].
    \end{equation}
    The above analysis leads to 
    \begin{equation}
        q \big(\vec{x}, \vec{y}\big) = \vec{\chi} \cdot \vec{f}(\vec{y}) - \vec{\beta} \cdot \vec{y}.
    \end{equation}
    With the concavity of $f_r^k (\cdot)$, the result $(ii)$ is immediate.
\end{proof}
%%%%%% Proof Sketch
% \begin{proof}
%     Due to the space limits, we only give the proof sketch here. The main technique is to find the vectorized representation of $\mathcal{Y}$ and $q(\cdot)$. 
%     A series of transformations lead to $\mathcal{Y}$ being $\big \{\vec{y} \mid \vec{0} \leq \vec{y} \leq \vec{a}, \mathbf{B} \vec{y} \leq \vec{c} \big\}$, i.e., 
%     a polyhedron. It is well known to be convex. Similarly, we have 
%     $q \big(\vec{x}, \vec{y}\big) = \vec{\chi} \cdot \vec{f}(\vec{y}) - \vec{\beta} \cdot \vec{y}$.
%     With the concavity of $f_r^k (\cdot)$, the result $(ii)$ is immediate.
% \end{proof}

As a result, the derivative of $q(\cdot)$ at time $t$ is 
\begin{equation}
    \frac{\partial q\big(\vec{x}(t), \vec{y}(t)\big)}{\partial y_{(l,r)}^k(t)} = 
    \left\{
        \begin{array}{ll}
            x_l(t) \Big( (f_r^k)'\big(y_{(l,r)}^k(t)\big) - \beta_k \Big) & k = k^* \\
            x_l(t) (f_r^k)'(y_{(l,r)}^k(t)) & \textrm{o.w.},
        \end{array}
    \right.
    \label{derive}
\end{equation}
where $k^*$ is defined by \eqref{k-star}.

\subsection{Online Gradient Ascent}\label{s3.2}
In this section, we give the design details of the OGA-based bipartite scheduling policy.
\begin{definition}
    \textsc{The OGA Policy}. For any feasible initial bipartite scheduling decision 
    $\vec{y}(1) \in \mathcal{Y}$, at each time $t \in \mathcal{T}$, the OGA policy gets $\vec{y}(t+1)$ in the direction of 
    ascending the gradient of $q\big( \vec{x}(t), \vec{y}(t) \big)$:
    \begin{equation}
        \vec{y}(t+1) = \Pi_{\mathcal{Y}} \Big( \vec{y}(t) + \eta_{t} \nabla q \big( \vec{x}(t), \vec{y}(t) \big) \Big),
    \end{equation}
    where $\eta_t$ is the step size, and
    \begin{flalign}
        \Pi_{\mathcal{Y}} (\vec{z}) = \argmin_{\hat{\vec{y}} \in \mathcal{Y}} \big\| \hat{\vec{y}} - \vec{z} \big\|_2^2
        \label{project}
    \end{flalign}
    is the Euclidean projection of $\vec{z}$ onto $\mathcal{Y}$. 
\end{definition}

To implement the projection \eqref{project} with low complexity, we propose \textsc{OgaSched}, which is a combination of the OGA policy and the following fast projection technique. Firstly, we introduce the Lagrangian of the projection \eqref{project} as
\begin{flalign}
    &L(\hat{\vec{y}}, \vec{\rho}, \vec{\mu}, \vec{\lambda}) 
    = \sum_{l \in \mathcal{L}} \sum_{r \in \mathcal{R}_l} \sum_{k \in \mathcal{K}} \Big( \hat{y}_{(l,r)}^k - z_{(l,r)}^k \Big)^2 \nonumber \\
    &+ \sum_{r \in \mathcal{R}} \sum_{k \in \mathcal{K}} \rho_r^k \Big(\sum_{l \in \mathcal{L}_r} \hat{y}_{(l,r)}^k - c_r^k \Big)  - \sum_{l \in \mathcal{L}} \sum_{r \in \mathcal{R}_l} \sum_{k \in \mathcal{K}} \lambda_{l,r}^k \hat{y}_{(l,r)}^k\nonumber \\
    &\qquad \qquad + \sum_{l \in \mathcal{L}} \sum_{r \in \mathcal{R}_l} \sum_{k \in \mathcal{K}} \mu_{l,r}^k \Big( \hat{y}_{(l,r)}^k - a_l^k \Big),
\end{flalign}
where $\vec{\rho}$ is the dual variable for \eqref{cons2}, $\vec{\mu}$ is the dual variable for $\vec{y}(t) \leq \vec{a}$, and $\vec{\lambda}$ is the dual variable for $\vec{y}(t) \geq \vec{0}$. Then, we can write the KKT conditions of the projection as
% \begin{flalign}
%     &2 \big( \hat{y}_{(l,r)}^k - z_{(l,r)}^k \big) + \rho_r^k - \lambda_{l,r}^k + \mu^k_{l,r} = 0 \label{kkt1}\\
%     &\qquad \sum_{l \in \mathcal{L}_r} \hat{y}_{(l,r)}^k = c_r^k \textrm{ \& } \rho_{r}^k > 0 \label{kkt2}\\
%     &\quad \qquad \hat{y}_{(l,r)}^k = a_l^k \textrm{ \& } \mu_{l,r}^k > 0 \label{kkt3}\\
%     &\quad \qquad \hat{y}_{(l,r)}^k = 0 \textrm{ \& } \lambda_{l,r}^k > 0
%     \label{kkt-equations}
% \end{flalign}
\begin{eqnarray}
    \left\{
        \begin{array}{l}
            2 \big( \hat{y}_{(l,r)}^k - z_{(l,r)}^k \big) + \rho_r^k - \lambda_{l,r}^k + \mu^k_{l,r} = 0 \\
            \sum_{l \in \mathcal{L}_r} \hat{y}_{(l,r)}^k = c_r^k \textrm{ \& } \rho_{r}^k > 0 \\
            \hat{y}_{(l,r)}^k = a_l^k \textrm{ \& } \mu_{l,r}^k > 0 \\
            \hat{y}_{(l,r)}^k = 0 \textrm{ \& } \lambda_{l,r}^k > 0
        \end{array}
    \right.
    \label{kkt}
\end{eqnarray}
for every $l,r,k$.

Our fast projection is implemented for each pair of $(r,k)$ in parallel. Specifically, for each $r \in \mathcal{R}$ and each $k \in \mathcal{K}$, we divide the ports $l \in \mathcal{L}$ into three disjoint sets:
\begin{eqnarray*}
    \left\{
        \begin{array}{l}
            \mathcal{B}^1_{rk} = \Big\{ l \in \mathcal{L}_r \mid \forall (l,r,k): \hat{y}_{(l,r)}^k = a_l^k \Big\} \\
            \mathcal{B}^2_{rk} = \Big\{ l \in \mathcal{L}_r \mid  \forall (l,r,k): \hat{y}_{(l,r)}^k = 0 \Big\} \\
            \mathcal{B}^3_{rk} = \Big\{  l \in \mathcal{L}_r \mid \forall (l,r,k): 2 \big( \hat{y}_{(l,r)}^k - z_{(l,r)}^k \big) + \rho_r^k = 0 \Big\},
        \end{array}
    \right.
\end{eqnarray*}
where
\begin{equation}
    \rho_r^k = \frac{2}{|\mathcal{B}_{rk}^3|} 
    \bigg( \sum_{l \in \mathcal{B}_{rk}^3} z_{(l,r)}^k - c_r^k + \sum_{l \in \mathcal{B}_{rk}^1} a_l^k \bigg), \forall r, k.
    \label{rho}
\end{equation}
The fast projection works by solving the equation system \eqref{kkt} iteratively. Specifically, for each pair of $(r, k)$, we sort the elements of $\vec{z}_{(:,r)}^k$ in descending order (step 7), and initialize $\mathcal{B}^1_{rk}$ and $\mathcal{B}^2_{rk}$ as $\varnothing$ while initializing $\mathcal{B}^3_{rk}$ as $\mathcal{L}_r$ (step 10 and 12). Then, we repeat a loop, in which we calculate $\rho_r^k$ with \eqref{rho}, and update the value of $\hat{y}_{(l,r)}^k$ for each port $l$ in $\mathcal{B}_{rk}^3$ (step 25). Since the elements of $\vec{z}_{(:,r)}^k$ are sorted from largest to smallest, if some $\hat{y}_{(l,r)}^k < 0$, we can derive that for all the $l' \in \mathcal{S}_{rk} := \big\{l, ..., |\mathcal{L}_r| \big\}$, we have $\hat{y}_{(l',r)}^k < 0$. Thus, the resource allocation for all the ports in $\mathcal{S}_{rk}$ is illegal, since $\hat{y}_{(l,r)}^k \geq 0$ must hold. As a result, we update the sets $\mathcal{B}_{rk}^2$ and $\mathcal{B}_{rk}^3$, and repeat the calculate loop again (step 29). The calculation loop stops when there are no illegal resource allocations, i.e., $\forall l \in \mathcal{L}_r$, we have $\hat{y}_{(l,r)}^k \geq 0$. In other words, $\mathcal{S}_{rk} = \varnothing$.  We call the calculation loop in step 18 $\sim$ step 30 the innner loop. The outer loop is the while loop defined in step 9. To exit the while loop, we need to guarantee that $\hat{y}_{(1,r)}^k \leq a_1^k$. Otherwise, the resource allocation is also illegal. Note that here we only need to check for $l=1$ since the elements in $\vec{z}_{(:,r)}^k$ are sorted.

The number of projections is linearly proportional to the size of the solution's dimensions, i.e., $\sum_{l \in \mathcal{L}} |\mathcal{R}_l| \times K$. Nevertheless, as we have mentioned, we can do the projections for different combinations of $r$ and $k$ in parallel because they are not interwoven. Thus, the time complexity of the fast projection is of $\mathcal{O} \big( |\mathcal{L}| \times \log ( K \sum_{l \in \mathcal{L}} |\mathcal{R}_l| ) \big)$ in each time slot, where the $\log(\cdot)$ operator comes from the sorting operation (step 7). The multiplier $|\mathcal{L}|$ outside $\log(\cdot)$ comes from the inner loop (step 19). In our experiments, the repeat loop's execution count is significantly less than the number of job types $|\mathcal{L}|$.

% \begin{figure}
%     \removelatexerror
%     \begin{algorithm}[H]
%         \label{algo}
%         \caption{\textsc{RepeatLoop}}
%         \Repeat{$\mathcal{S}_{rk} = \varnothing$}
%         {
%             Calculate $\rho_r^k$ with \eqref{rho} \\
%             \For{$l \in \mathcal{L}_r$}
%             {
%                 \uIf{$l \in \mathcal{B}_{rk}^1$}
%                 {
%                     $\hat{y}_{(l,r)}^k \leftarrow a_l^k$\\
%                 }\uElseIf{$l \in \mathcal{B}_{rk}^3$}
%                 {
%                     $\hat{y}_{(l,r)}^k \leftarrow z_{(l,r)}^k (t+1) - \rho_r^k/2$\\
%                     \If{$\hat{y}_{(l,r)}^k < 0$}
%                     {
%                         $\mathcal{S}_{rk} \leftarrow \{ l, l+1, ..., |\mathcal{L}_r|\}$\\
%                         \textbf{break}
%                     }
%                 }
%             }
%             $B_{rk}^3 \leftarrow B_{rk}^3 \backslash \mathcal{S}_{rk}, B_{rk}^2 \leftarrow \mathcal{B}_{rk}^2 \cup \mathcal{S}_{rk}$\\
%         }
%     \end{algorithm}
% \end{figure}

\subsection{Regret Analysis}\label{s3.3}
In this section, we discuss the regret of \textsc{OgaSched}. The main result is summarized in Theorem \ref{theo1}.
\begin{theorem}
    \textsc{Regret Upper Bound}. With a nice setup, the regret of \textsc{OgaSched} is upper bounded by
    \begin{flalign}
        R_T^{\textrm{\textsc{OgaSched}}} &\leq \sqrt{2 T \sum_{k \in \mathcal{K}} \sum_{r \in \mathcal{R}} \bar{a}^k c_r^k} \nonumber \\
        &\times \sqrt{ \sum_{l \in \mathcal{L}} \sum_{r \in \mathcal{R}_l} \Big( (\beta^*)^2 +K  (\varpi_r^*)^2 \Big)},
    \end{flalign}
    where $\bar{a}^k := \max_{l \in \mathcal{L}} a_l^k$, $\beta^* := \max_{k \in \mathcal{K}} \beta_k$, and 
    $\varpi_r^* := \max_{k \in \mathcal{K}} \varpi_r^k$.
    \label{theo1}
\end{theorem}
\begin{proof}
    The result is based on the non-expansiveness property of Euclidean projection and the concavity of $\{f_r^k{(\cdot)}\}_{r,k}$. 
    Our proof has two parts. {\color{black}The first part gives the general form of the upper bound, which is similar to Theorem 2.13 in \cite{orabona2019modern} and Theorem 3 in \cite{23}. Meanwhile, the second part gives the specific upper bounds of involved variables.}

    \begin{figure}
        \removelatexerror
        \begin{algorithm}[H]
            \label{oga}
            \caption{\textsc{OgaSched}}
            \KwIn{\color{black}Graph $\mathcal{G}$, requirements $\vec{a}$, capacities $\vec{c}$, and the decay $\lambda$}
            \KwOut{Scheduling decisions $\{\vec{y}(t)\}_{t \in \mathcal{T}}$}
            Initialize $\vec{y}(1) \in \mathcal{Y}$ and $\eta_0$ \\
            \For{$t$ from $1$ to $T$}
            {
                Observe the job arrival status $\vec{x}(t)$ \\
                Calculate the gradient $\nabla q \big( \vec{x}(t), \vec{y}(t) \big)$ with \eqref{derive} \\
                {\color{black}$\vec{z}(t+1) \leftarrow \vec{y}(t) + \eta_{t} \nabla q \big( \vec{x}(t), \vec{y}(t) \big)$} \\
                \ForEach(\textbf{in parallel}){$(r, k)$ in $\texttt{zip}(\mathcal{R}, \mathcal{K})$}
                {
                    Sort the elements of $\vec{z}_{(:, r)}^k (t+1)$ in descending order \\
                    % : $z_{(1,r)}^k (t + 1) \geq ... \geq z_{(|\mathcal{L}|, r)}^k (t + 1)$ \\
    
                    {\color{black}$\textit{initialized} \leftarrow \texttt{False}$} \\
    
                    \While{\normalfont \texttt{True}}
                    {
                        $\mathcal{B}_{rk}^2 \leftarrow \varnothing$\\
                        \If{\color{black}{\normalfont \textbf{not}} \textit{initialized}}
                        {
                            $\mathcal{B}_{rk}^1 \leftarrow \varnothing, 
                            \mathcal{B}_{rk}^3 \leftarrow \mathcal{L}_r,
                            \hat{\vec{y}} \leftarrow \vec{0}$ \\
                            {\color{black}$\textit{initialized} \leftarrow \texttt{True}$}\\
                        }\Else{
                            \If{$\hat{y}_{(1,r)}^k > a_1^k$}
                            {
                                $\mathcal{B}_{rk}^1 \leftarrow \{ 1 \}, 
                                \mathcal{B}_{rk}^3 \leftarrow \mathcal{L}_r \backslash \{ 1 \}$ \\
                            }\Else
                            {
                                \textbf{break}\\
                            }
                        }
                        \Repeat{$\mathcal{S}_{rk} = \varnothing$}
                        {
                            Calculate $\rho_r^k$ with \eqref{rho} \\
                            \For{$l \in \mathcal{L}_r$}
                            {
                                \If{$l \in \mathcal{B}_{rk}^1$}
                                {
                                    $\hat{y}_{(l,r)}^k \leftarrow a_l^k$\\
                                }\ElseIf{$l \in \mathcal{B}_{rk}^3$}
                                {
                                    $\hat{y}_{(l,r)}^k \leftarrow z_{(l,r)}^k (t+1) - \rho_r^k/2$\\
                                    \If{$\hat{y}_{(l,r)}^k < 0$}
                                    {
                                        $\mathcal{S}_{rk} \leftarrow \Big\{ l, l+1, ..., |\mathcal{L}_r| \Big\}$\\
                                        \textbf{break}
                                    }
                                }
                            }
                            \tcp{Update then re-calculate}
                            $\mathcal{B}_{rk}^2 \leftarrow \mathcal{B}_{rk}^2 \cup \mathcal{S}_{rk}, B_{rk}^3 \leftarrow B_{rk}^3 \backslash \mathcal{S}_{rk}$\\
                        }
                    }
                }
                $\vec{y}(t+1) \leftarrow \hat{\vec{y}}$ \\
                {\color{black}$\eta_{t+1} \leftarrow \lambda \eta_{t}$} \tcp{Update learning rate}
            }
            \Return{the sequence of decisions $\{\vec{y}(t)\}_{t \in \mathcal{T}}$}
        \end{algorithm}
    \end{figure}

    At each time $t > 1$, for the $\vec{y}(t)$ yielded by \textsc{OgaSched}, we have 
    \begin{flalign}
        \| \vec{y}(t) - \vec{y}^* \|^2 
        &= \big\| \Pi_{\mathcal{Y}} \big( \vec{y}(t-1) + \eta_{t} \nabla q(t-1) \big) - \vec{y}^* \big\|^2 \nonumber \\
        &\overset{(i)}{\leq} \| \vec{y}(t-1) - \vec{y}^* \|^2 + \eta_t^2 \| \nabla q(t-1) \|^2 \nonumber \\
        &+ 2 \eta_t \nabla q\big(\vec{y}(t-1)\big)^{\textrm{T}} \big(\vec{y}(t-1) - \vec{y}^*\big),
        \label{prove-1}
    \end{flalign}
    where $\nabla q\big(\vec{y}(t-1)\big)$ is a shorthand for $\nabla q \big( \vec{x}(t-1), \vec{y}(t-1) \big)$. 
    $(i)$ is because the non-expansiveness property of the Euclidean projection. 
    By moving $\| \vec{y}(t-1) - \vec{y}^* \|^2$ to the LHS of 
    \eqref{prove-1} and summing the inequality telescopically over $\mathcal{T}$, 
    we have 
    \begin{flalign}
        &\qquad \qquad \qquad \sum_{t=2}^{T+1} \nabla q (\vec{y}(t-1))^{\textrm{T}} \big(\vec{y}^* - \vec{y}(t-1)\big) \nonumber \\
        &\overset{(i)}{\leq} \frac{\eta \sum_{t=1}^T \| \nabla q(\vec{y}(t)) \|^2}{2} + \frac{\| \vec{y}(1) - \vec{y}^* \|^2 - \| \vec{y}(T) - \vec{y}^* \|^2}{2 \eta} \nonumber \\
        &\qquad \qquad \qquad \overset{(ii)}{\leq} \frac{\eta T (\max \| \nabla q \|)^2}{2} + \frac{\textrm{\textit{diam}}(\mathcal{Y})^2}{2 \eta}.
    \end{flalign}
    Inequality $(i)$ is because $\forall t \in \mathcal{T}$ we set $\eta_t \equiv \eta$. In $(ii)$, we use the fact that 
    $\| \vec{y}(T) - \vec{y}^* \| \geq 0$. In \eqref{prove-1}, $\max \| \nabla q \|$ is the maximum Euclidean norm of the gradient of 
    $q (\vec{x}(t), \vec{y}(t))$ over every possible $\vec{y}(t)$, and $\textrm{\textit{diam}}(\mathcal{Y})$ is the largest Euclidean 
    distance between any two elements of $\mathcal{Y}$. Because $q(\cdot)$ is a concave function of $\vec{y}(t)$, we have
    \begin{flalign}
        R_T^{\textrm{\textsc{OgaSched}}} 
        &= \sup_{\forall \{\vec{x}(t)\}_1^T} \sum_{t=1}^T \Big( q \big(\vec{x}(t), \vec{y}^*\big) - q\big(\vec{x}(t), \vec{y}(t)\big) \Big) \nonumber \\
        &\leq \sup_{\forall \{\vec{x}(t)\}_1^T} \sum_{t=1}^T \nabla q \big(\vec{y}(t)\big)^\textrm{T} \big(\vec{y}^* - \vec{y}(t)\big) \quad \rhd \eqref{prove-1} \nonumber \\
        &\leq \frac{\textrm{\textit{diam}}(\mathcal{Y})^2}{2 \eta} + \frac{\eta T (\max \| \nabla q \|)^2}{2}. 
        \label{prove-2}
    \end{flalign}

    In the following, we give the upper bound of $\max \| \nabla q \|$ and $\textrm{\textit{diam}}(\mathcal{Y})$, respectively. 

    1) \textit{The upper bound of} $\max \| \nabla q \|$. With the result of \eqref{derive}, we have 
    \begin{flalign}
        \|\nabla q\|^2 
        &= \sum_{l \in \mathcal{L}} \sum_{r \in \mathcal{R}_l} \bigg[ x_l(t)^2 \Big( (f_r^{k^*})' \big(y_{(l,r)}^{k^*}(t)\big) - \beta_{k^*} \Big)^2 \bigg] \nonumber \\
        &+ \sum_{l \in \mathcal{L}} \sum_{r \in \mathcal{R}_l} \sum_{k \neq k^*} x_l(t)^2 (f_r^k)'\Big(y_{(l,r)}^k(t)\Big)^2 \nonumber \\
        &= \sum_{l \in \mathcal{L}} \sum_{r \in \mathcal{R}_l} x_l(t)^2 \bigg[\sum_{k \in \mathcal{K}} \Big( (f_r^{k})' \big(y_{(l,r)}^{k}(t)\big) \Big)^2 \nonumber \\
        &- 2 \beta_{k^*} (f_r^{k^*})' \big(y_{(l,r)}^{k^*}(t)\big) \bigg] + \sum_{l \in \mathcal{L}} \sum_{r \in \mathcal{R}_l} x_l(t)^2 \beta_{k^*}^2.
        \label{prove-5}
    \end{flalign}
    where $k^*$ is defined in \eqref{k-star}. The second part of \eqref{prove-5} can be upper bounded by
    \begin{equation}
        \sum_{l \in \mathcal{L}} \sum_{r \in \mathcal{R}_l} x_l(t)^2 \beta_{k^*}^2 \leq \sum_{l \in \mathcal{L}} \sum_{r \in \mathcal{R}_l} (\beta^*)^2,
        \label{prove-6}
    \end{equation}
    where $\beta^* = \max_{k \in \mathcal{K}} \beta_k$. {\color{black}If $\mathcal{G}$ is right $d$-regular, the bound reduces to 
    $d |\mathcal{R}| (\beta^*)^2$.} For the first part of \eqref{prove-5}, we use $(f_r^{k^*})'$ to replace $(f_r^{k^*})' \big(y_{(l,r)}^{k^*}(t)\big)$ 
    for simplification. Then we have
    \begin{flalign*}
        &\quad \sum_{l \in \mathcal{L}} 
        \sum_{r \in \mathcal{R}_l} x_l(t)^2 \Big[ \sum_{k \in \mathcal{K}} \big( (f_r^{k})' \big)^2 - 2 \beta_{k^*} (f_r^{k^*})'\Big] \nonumber \\
        &\leq \underbrace{\sum_{l \in \mathcal{L}} \sum_{r \in \mathcal{R}_l} \sum_{k \neq k^*} \big( (f_r^{k})' \big)^2}_{\textrm{\textsc{Part-A}}} + 
        \underbrace{\sum_{l \in \mathcal{L}} \sum_{r \in \mathcal{R}_l} (f_r^{k^*})' \big( (f_r^{k^*})' - 2 \beta_{k^*} \big)}_{\textrm{\textsc{Part-B}}}. 
    \end{flalign*}
    For \textsc{Part-A} we have
    \begin{flalign}
        \textrm{\textsc{Part-A}} \leq (K-1) \sum_{l \in \mathcal{L}} \sum_{r \in \mathcal{R}_l} (\varpi_r^*)^2,
    \end{flalign}
     where $\varpi_r^* = \max_{k \in \mathcal{K}} \varpi_r^k$.
    {\color{black}If $\mathcal{G}$ is right $d$-regular, the bound reduces to $d |\mathcal{R}| (K-1) (\varpi_r^*)^2$}. To analyze the upper bound of \textsc{Part-B}, 
    we need to partition the computing instances into two disjoint sets:
    \begin{flalign}
        \mathcal{R}_1 &= \Big\{ r \in \mathcal{R}: \varpi_r^{k^*} \leq 2 \beta_{k^*} \Big\} \nonumber \\
        \mathcal{R}_2 &= \Big\{ r \in \mathcal{R}: \varpi_r^{k^*} > 2 \beta_{k^*} \Big\}. \nonumber
    \end{flalign}
    For each $r \in \mathcal{R}_1$, the maximum of $(f_r^{k^*})' \big( (f_r^{k^*})' - 2 \beta_{k^*} \big)$ is $0$ since $(f_r^{k^*})' \geq 0$ holds. 
   For each $r \in \mathcal{R}_2$, the maximum is $(\varpi_r^{k^*})^2 - 2 \beta_{k^*} \varpi_r^{k^*}$. Thus, 
    \begin{flalign}
        \textrm{\textsc{Part-B}} \leq \sum_{l \in \mathcal{L}} \sum_{r \in \mathcal{R}_l \cap \mathcal{R}_2} \Big( (\varpi_r^{k^*})^2 - 2 \beta_{k^*} \varpi_r^{k^*} \Big).
        \label{p9}
    \end{flalign}
    Recall that in \eqref{p9} $\mathcal{R}_l$ is the set of computing instances that connects to port $l$. 
    Because $\beta_k \in [0, 1]$ holds for each $k \in \mathcal{K}$, $\forall l \in \mathcal{L}, r \in \mathcal{R}_l \cap \mathcal{R}_2$, we have 
    \begin{flalign}
        (\varpi_r^{k^*})^2 - 2 \beta_{k^*} \varpi_r^{k^*} \leq 
        (\varpi_r^*)^2 - 2 \beta_{k^*} \varpi_r^* \leq  (\varpi_r^*)^2,
        \label{p10}
    \end{flalign}
    Finally, we can get 
    \begin{equation}
        \|\nabla q\|^2 \leq \sum_{l \in \mathcal{L}} \sum_{r \in \mathcal{R}_l} \Big( (\beta^*)^2 +K  (\varpi_r^*)^2 \Big).
        \label{p11}
    \end{equation}
    For the upper bound in \eqref{p11}, all the computing instances $r \in \mathcal{R}_l$ fall into the set $\mathcal{R}_2$.

    2) \textit{The upper bound of} $\textrm{\textit{diam}}(\mathcal{Y})$. By definition we have 
    \begin{flalign}
        \textrm{\textit{diam}}(\mathcal{Y}) = \sup_{\vec{y}, \vec{z} \in \mathcal{Y}} \| \vec{y} - \vec{z} \|.
    \end{flalign}
    To find the upper bound of $\| \vec{y} - \vec{z} \|$, we can get 
    \begin{flalign}
        \| \vec{y} - \vec{z} \|^2 &= \sum_{l \in \mathcal{L}} \sum_{r \in \mathcal{R}_l} \sum_{k \in \mathcal{K}} \Big( y_{(l,r)}^k - z_{(l,r)}^k \Big)^2 \nonumber \\
        &\overset{(i)}{\leq} \sum_{l \in \mathcal{L}} \sum_{r \in \mathcal{R}_l} \sum_{k \in \mathcal{K}} \big| y_{(l,r)}^k - z_{(l,r)}^k  \big| \cdot a_l^k \nonumber \\
        &\leq \sum_{l \in \mathcal{L}} \sum_{r \in \mathcal{R}_l} \sum_{k \in \mathcal{K}} a_l^k \Big( y_{(l,r)}^k + z_{(l,r)}^k \Big) \nonumber \\
        &\leq \sum_{k \in \mathcal{K}} \bar{a}^k \sum_{r \in \mathcal{R}} \Big( \sum_{l \in \mathcal{L}_r} y_{(l,r)}^k + \sum_{l \in \mathcal{L}_r} z_{(l,r)}^k \Big) \nonumber \\
        &\overset{(ii)}{\leq} 2 \sum_{k \in \mathcal{K}} \bar{a}^k \sum_{r \in \mathcal{R}} c_r^k,
        \label{prove-3}
    \end{flalign}
    where $\bar{a}^k = \max_{l \in \mathcal{L}} a_l^k$. In \eqref{prove-3}, $(i)$ is because the constraint \eqref{cons1}. In $(ii)$, we use the capacity constraint \eqref{cons2}. 
    As a result, we have 
    \begin{equation}
        \textrm{\textit{diam}}(\mathcal{Y}) \leq \sqrt{2 \sum_{k \in \mathcal{K}} \bar{a}^k \sum_{r \in \mathcal{R}} c_r^k}.
        \label{prove-4}
    \end{equation}

    Combing the result \eqref{p11} and \eqref{prove-4}, and set $\eta$ as $\frac{\textrm{\textit{diam}}(\mathcal{Y})}{\|\nabla q\| \sqrt{T}}$, we finally 
    get the result.
\end{proof}
{\color{black}The theorem shows that the suboptimality gap between \textsc{OgaSched} and the offline optimal is of 
$\Theta (\mathcal{H}_{\mathcal{G}} \times \sqrt{T})$, where
\begin{flalign}
    \mathcal{H}_{\mathcal{G}} := \sqrt{2 \sum_{k \in \mathcal{K}} \sum_{r \in \mathcal{R}} \bar{a}^k c_r^k} \times \sqrt{ \sum_{l \in \mathcal{L}} \sum_{r \in \mathcal{R}_l} \Big( (\beta^*)^2 + K (\varpi_r^*)^2 \Big)}
    \label{hg}
\end{flalign}
is a factor characterized the scale of the bipartite graph $\mathcal{G}$.} In addition, we can find that the regret grows sublinearly with the number of job types $|\mathcal{L}|$. To the best of our knowledge, this is the best regret for the online bipartite scheduling problem with non-linear rewards. The proof also indicates that, to achieve a not-too-bad cumulative reward, at each time $t$, the learning rate $\eta_t$ should be set as 
\begin{equation}
    \eta_t = \frac{\textrm{\textit{diam}}(\mathcal{Y})}{\| \nabla q(\vec{x} (t), \vec{y}(t)) \| \sqrt{T}}.
    \label{set-eta}
\end{equation}

\subsection{Extending to Multiple Job Arrivals}\label{s3.4}
\textsc{OgaSched} can be applied to the scenarios where multiple jobs are yielded from each port in each time slot. 
In this case, the job arrival status $\vec{x}(t)$ is re-formulated as $\vec{x} (t) = \big[x_l(t)\big]_{l \in \mathcal{L}} \in \mathbb{N}^{|\mathcal{L}|}$,
where $x_l(t)$ indicates the number of jobs arrive at port $l$ at time $t$. Further, the scheduling decisions at time $t$ is re-formulated as
\begin{equation*}
    \vec{y}(t) = \Big[ y_{(l,r)}^{j,k} \Big]_{l,j,r,k} \in 
    \mathbb{R}^{\sum_{l \in \mathcal{L}} J_l \times |\mathcal{R}_l| \times K },
\end{equation*}
where $J_l$ is the maximum number of the type-$l$ jobs arrive during each time slot, i.e. $J_l = \max_{t \in \mathcal{T}} x_l(t)$. Correspondingly, 
the port-$l$ reward is re-formulated as
\begin{flalign*}
    q_l\big(\vec{x}(t), \vec{y}(t)\big) 
    = \sum_{j = 1}^{J_l} \mathds{1} \{ j \leq x_l(t) \} \bigg[ &\sum_{k \in \mathcal{K}} f_k \Big( \sum_{r \in \mathcal{R}_l} y_{(l,r)}^{j,k}(t)\Big) - \nonumber \\
    &\max_{k \in \mathcal{K}} \Big\{ \beta_k \sum_{r \in \mathcal{R}_l} y_{(l,r)}^{j,k}(t) \Big\} \bigg],
\end{flalign*}
where $\mathds{1} \{ p \}$ is the indicator function: $\mathds{1} \{ p \}$ is $1$ if the predicate $p$ is true, otherwise $0$. 
The new formulated problem can be solved by native \textsc{OgaSched} after transformations.

\subsection{Extending to Gang Scheduling}\label{s3.5}
\textsc{OgaSched} can be extended to the Gang Scheduling scenarios, where the scheduling decisions for the task instances of 
a job follows the \textsc{All-or-Nothing} property. In other words, only when \textit{all} tasks\footnote{In practice, not all tasks 
of a job need to be scheduled. In Kubernetes, the job submitter can specify the minimum number of tasks that 
must be scheduled successfully. In the following, we use $m_l(t)$ to represent the minimum number of tasks that should be 
scheduled at time $t$ of the type-$l$ job.} of a job are successfully scheduled, the job could be launched. 

In the following, we show briefly how Gang Scheduling can be modeled. To start with, for each job type $l \in \mathcal{L}$, 
we denote the corresponding set of task components by $\mathcal{Q}_l$ and indexed by $q$. Correspondingly, the job requests 
$\vec{a}_l$ is redefined as $$\vec{a}_l = \big[ a_l^{q,k} \big]_{l,q,k} \in \mathbb{R}^{\sum_{l \in \mathcal{L}} |\mathcal{Q}_l| \times K}_{\geq 0}.$$
% \begin{equation*}
%     \vec{a}_l = \Big[ a_l^{q,k} \Big]_{l,q,k} \in \mathbb{R}^{\sum_{l \in \mathcal{L}} |\mathcal{Q}_l| \times K}.
% \end{equation*}
Similarly, we redefine the scheduling decisions at time $t$ as $$\vec{y}(t) = \Big[ y_{(l,r)}^{q,k} \Big]_{l,q,r,k} \in 
\mathbb{R}^{\sum_{l \in \mathcal{L}} |\mathcal{Q}_l| \times |\mathcal{R}_l| \times K }_{\geq 0}.$$
As a result, the solution space $\mathcal{Y}$ turns to
\begin{flalign*}
    \mathcal{Y} = \Big\{ y_{(l,r)}^{q,k} 
    &\mid \sum_{q \in \mathcal{Q}_l} \mathds{1} \big\{ \sum_{r \in \mathcal{R}_l} \sum_{k \in \mathcal{K}} y_{(l,r)}^{q,k} > 0  \big\} \geq m_l(t), \forall l, \nonumber \\
    &\quad 0 \leq y_{(l,r)}^{q,k}(t) \leq a_l^{q,k}, \forall l,r,q,k,t, \nonumber\\
    &\quad \sum_{l \in \mathcal{L}_r} \sum_{q \in \mathcal{Q}_l} y_{(l,r)}^{q,k} (t) \leq c_r^k, \forall r, k, t\Big\}, 
\end{flalign*}
where in the first inequality, $m_l(t)$ is the minimum number of task components that should be scheduled at time $t$ of type-$l$ job. 
The port-$l$ reward at time $t$ is re-formulated as 
\begin{flalign*}
    q_l\big(\vec{x}(t), \vec{y}(t)\big) 
    = x_l(t) \bigg[ &\sum_{k \in \mathcal{K}} f_k \Big(\sum_{q \in \mathcal{Q}_l} \sum_{r \in \mathcal{R}_l} y_{(l,r)}^{q,k}(t)\Big) - \nonumber \\
    &\max_{k \in \mathcal{K}} \Big\{ \beta_k \sum_{q \in \mathcal{Q}_l} \sum_{r \in \mathcal{R}_l} y_{(l,r)}^{q,k}(t) \Big\} \bigg].
\end{flalign*}

The new formulated problem is more difficult because $\mathcal{Y}$ is no longer a convex set and $q_l\big(\vec{x}(t), \vec{y}(t)\big)$ 
is not differentiable everywhere. Nevertheless, we can still develop a similar online scheduling algorithm with the subgradient ascent and mirror ascent 
related techniques which retains a sublinear regret. The design detail is omitted due to space limits.
% ======================================================================================================================================================

\section{Experimental Results}\label{s4}

\begin{figure*} 
    \centering 
    \subfigure[Average rewards.]{ 
        \label{2-a}
        \includegraphics[height=1.7in]{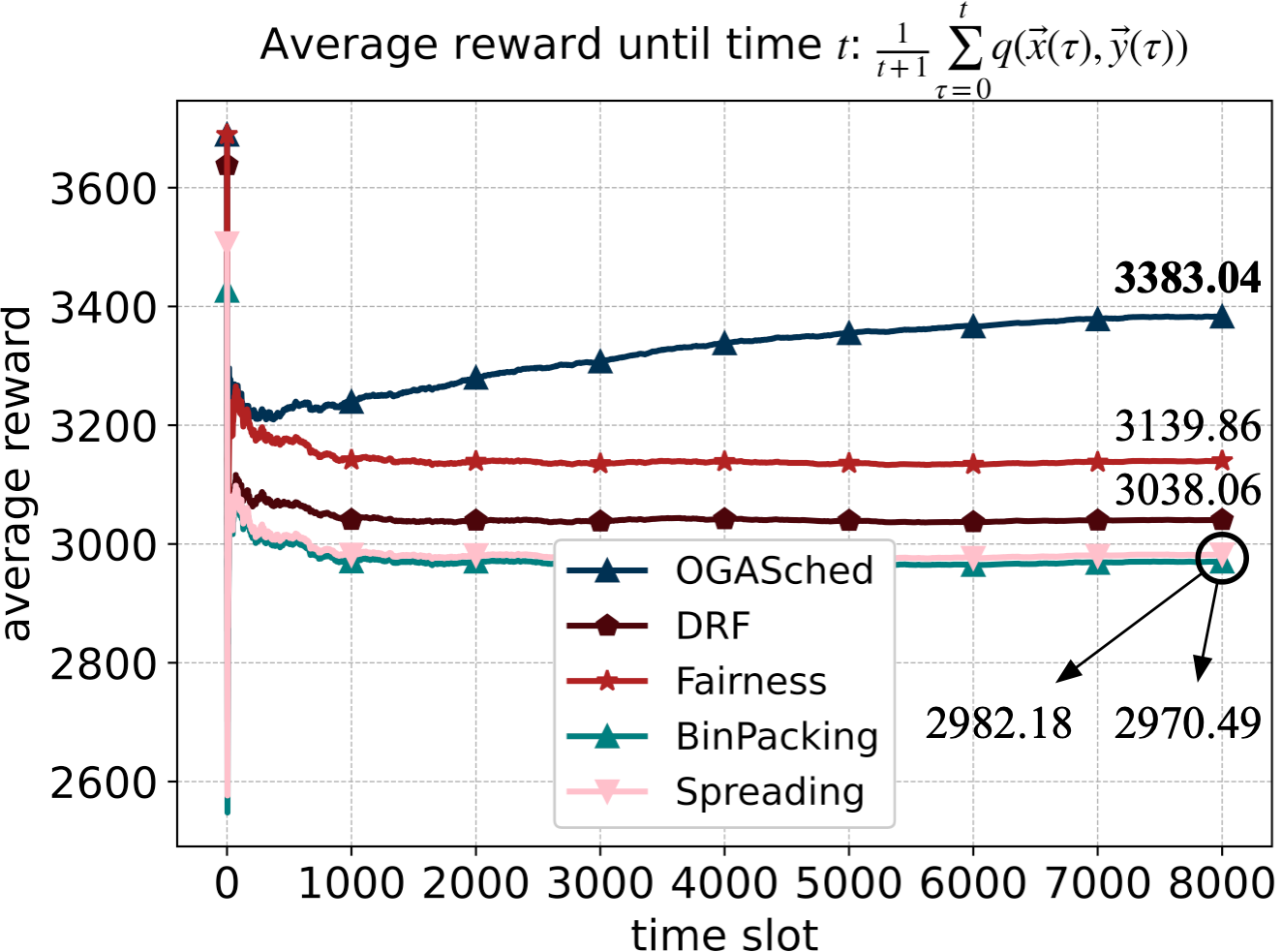}
    }
    \subfigure[Accumulative rewards.]{ 
        \label{2-b}
        \includegraphics[height=1.7in]{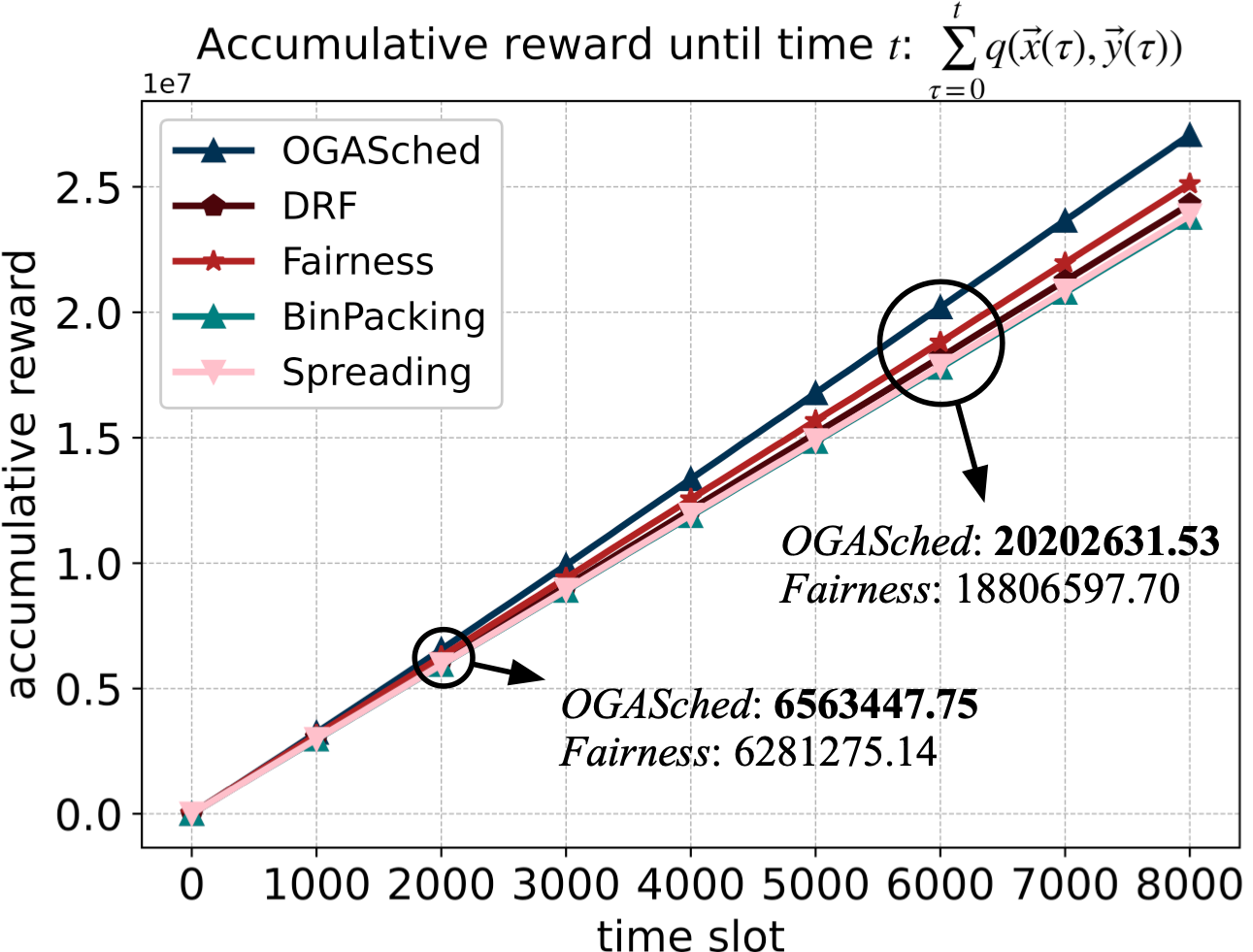}
    }
    \subfigure[Ratios between \textsc{OgaSched} and baselines.]{ 
        \label{2-c}
        \includegraphics[height=1.7in]{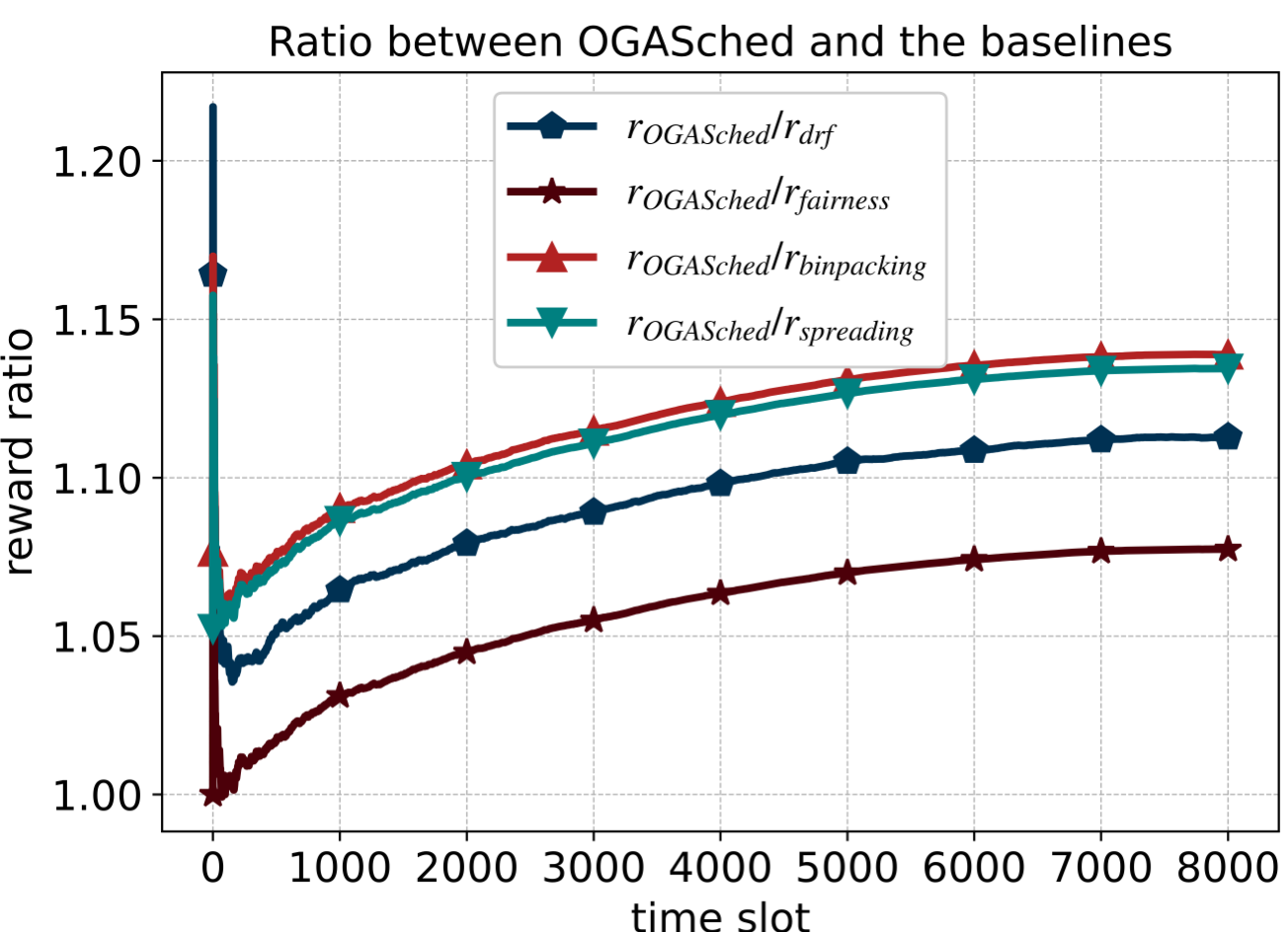}
    }
    \caption{Performance verification of \textsc{OgaSched}. It takes one hour for \textsc{OgaSched} to finish when $T = 8000$, $\beta \in [0.4, 0.6]$, and contention level is $11$.}
    \label{fig2}
\end{figure*}

In this section, we conduct extensive experiments to validate the performance of \textsc{OgaSched}. Based on the Alibaba cluster trace datasets \cite{trace}, we first examine the theoretically guaranteed superiority of \textsc{OgaSched} against several baselines on the cumulative and average rewards. Then, we analyze the generality and robustness of it under different cluster settings. At last, we validate the efficacy of \textsc{OgaSched} in large-scale scenarios. The trace-driven simulation is conducted on a server with 48 Intel Xeon Silver 4214 CPUs, 256 GB memory, and 2 Tesla P100 GPUs.

\textit{Traces}. We hybrid the traces from cluster-trace-v2018 and cluster-trace-gpu-v2020 of the Alibaba Cluster Trace Program. 
Specifically, we leverage the specifications of the machines, the arrival patterns, and the resource requirements of different kinds of jobs to generate our simulation environment. 

\textit{Baselines}. The following widely used baselines are implemented to make comparisons with \textsc{OgaSched}.
\begin{itemize}
    \item \textsc{DRF} \cite{DRF}. It is adopted by YARN \cite{yarn} and Mesos \cite{mesos}. In our scenario, \textsc{DRF} allocates resources to ports that yield jobs in the ascending order of their dominant resource shares. The dominant share $s_l$ of port $l$ is calculated as $s_l = \max_{k \in \mathcal{K}} \{ a_l^k / \sum_{r \in \mathcal{R}_l} c_r^k \}$. 
    
    \item \textsc{Fairness}. We implement \textsc{Fairness} in this way: at each time $t$, we allocate the type-$k$ resource of each node $r$ to each port $l$ that yield a job according to the job's share $a_l^k / \sum_{l \in \mathcal{L}_r} a_l^k$. 
    
    \item \textsc{BinPacking}. It is optional in Kubernetes with the name of \textsc{MostAllocated} strategy and supported in Volcano as a configurable plugin \cite{volcano}. Specifically, it scores the computing instances based on the utilization of resources, favoring the ones with higher allocation. 
    
    \item \textsc{Spreading}. It is similar to \textsc{BinPacking} in procedures but with an opposite favor. The nodes with lower utilizations 
    of resources have higher scores.
\end{itemize}

\textit{Default Settings}. In default settings, our simulation environment has 128 computing instances, each equipped with 6 types of 
resources (CPUs, MEM, GPUs, NPUs, TPUs, and FPGAs), and 10 job types of different resource requirements. 
Large-scale validations will be demonstrated in Sec. \ref{s4.3}. The computing instances and jobs are carefully selected from the trace to reflect heterogenity. We support 4 types of utilities:
\begin{flalign}
    f_r^k (y) = \left\{
        \begin{array}{ll}
            \alpha y & \textrm{\textit{linear}} \\
            \alpha \ln(y + 1) & \textrm{\textit{log}} \\
            \alpha^{-1} - (y + \alpha)^{-1} & \textrm{\textit{reciprocal}} \\
            \alpha \sqrt{y + 1} - \alpha & \textrm{\textit{poly}},
        \end{array}
    \right.
\end{flalign}

The default settings of main parameters are listed in Tab. \ref{tab2}. In this table, the initial learning rate and the decay are used to tune the learning rate at each time $t$ around the value \eqref{set-eta}. Job arrival probability $\rho$ is adopted to adjust the job arrival status with Bernoulli Distributions. This parameter is applied based on the actual arrival patterns from the trace to increase stochasticity. {\color{black}The contention level, located at the last cell of this table, is designed to tune the level of resource contention. The larger this value, the more fierce the contention. It is a multiplier to the resource requirements of jobs.} The effect of it will be analyzed in detail in Sec. \ref{s4.2}. 

\begin{table}[h]   
    \vspace{-0.15cm}
    \begin{center}
    \caption{\label{tab2}Default parameter settings.}   
    \begin{tabular}{c|c|c|c}    
        \toprule
        {\textsc{Parameter}} & {\textsc{Value}} & {\textsc{Parameter}} & {\textsc{Value}}\\[+0.1mm]
        \midrule
        job types num. $|\mathcal{L}|$ & $10$ & node num. $|\mathcal{R}|$ & $128$\\[+0.7mm]
        device type num. $K$ & $6$ & time slot num. $T$ & $2000$\\[+0.7mm]
        range of $\alpha$ & $[1.0, 1.5]$ & range of $\beta$ & $[0.3, 0.5]$\\[+0.7mm]
        initial learning rate $\eta_0$ & $25$ & {\color{black}decay $\lambda$} & $0.9999$ \\[+0.7mm]
        job arrival prob. $\rho$ & $0.7$ & contention level & $10$ \\[+0.7mm]
        \bottomrule   
    \end{tabular}  
    \end{center}
    \vspace{-0.15cm}
\end{table}

{\color{black}Note that in Sec. \ref{s4.1}, the time slot length $T$ is set as $8000$. For the left experiments, the time slot length is $2000$, unless otherwise stated.}

\subsection{Performance Verification}\label{s4.1}    
In this section, we compare the performance of \textsc{OgaSched} with the baselines in terms of the achieved cumulative and average rewards. 

\begin{figure*} 
    \centering 
    \subfigure[Accumulative rewards vs. $|\mathcal{R}|$.]{ 
        \label{4-a}
        \includegraphics[height=1.5in]{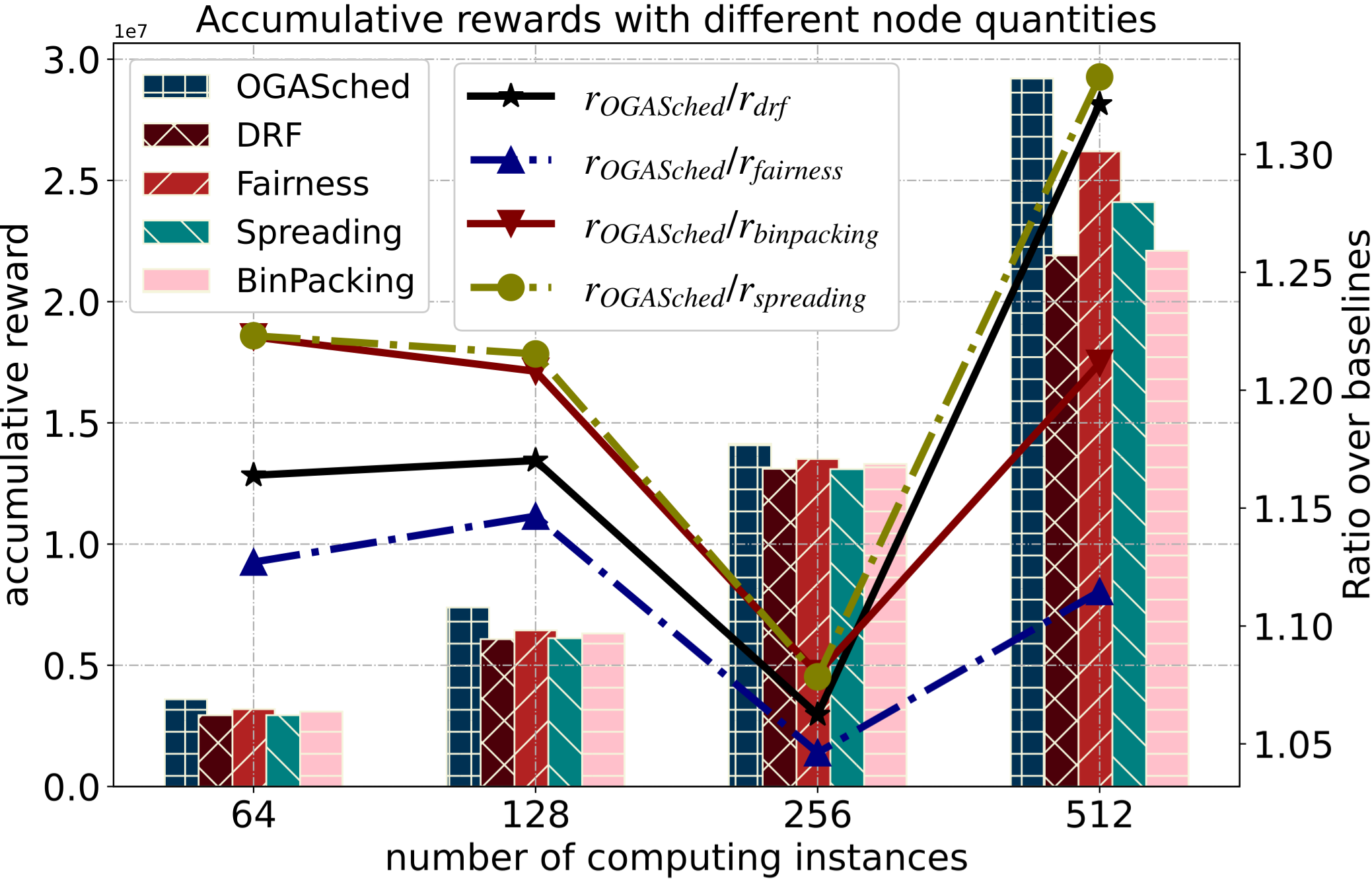}
    }
    \subfigure[Accumulative rewards vs. $|\mathcal{L}|$.]{ 
        \label{4-b}
        \includegraphics[height=1.5in]{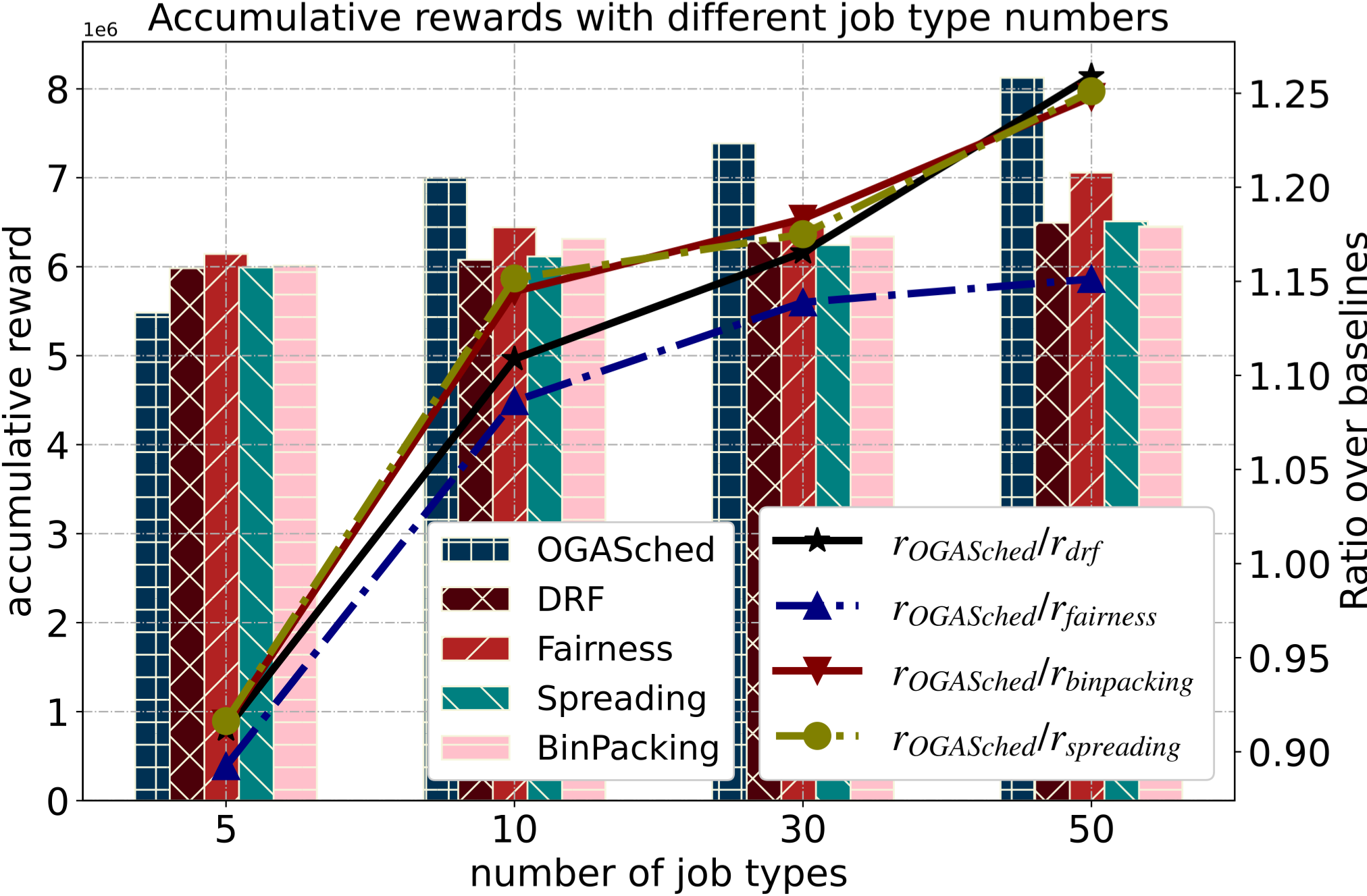}
    }
    \subfigure[Accumulative rewards vs. contention.]{ 
        \label{4-c}
        \includegraphics[height=1.5in]{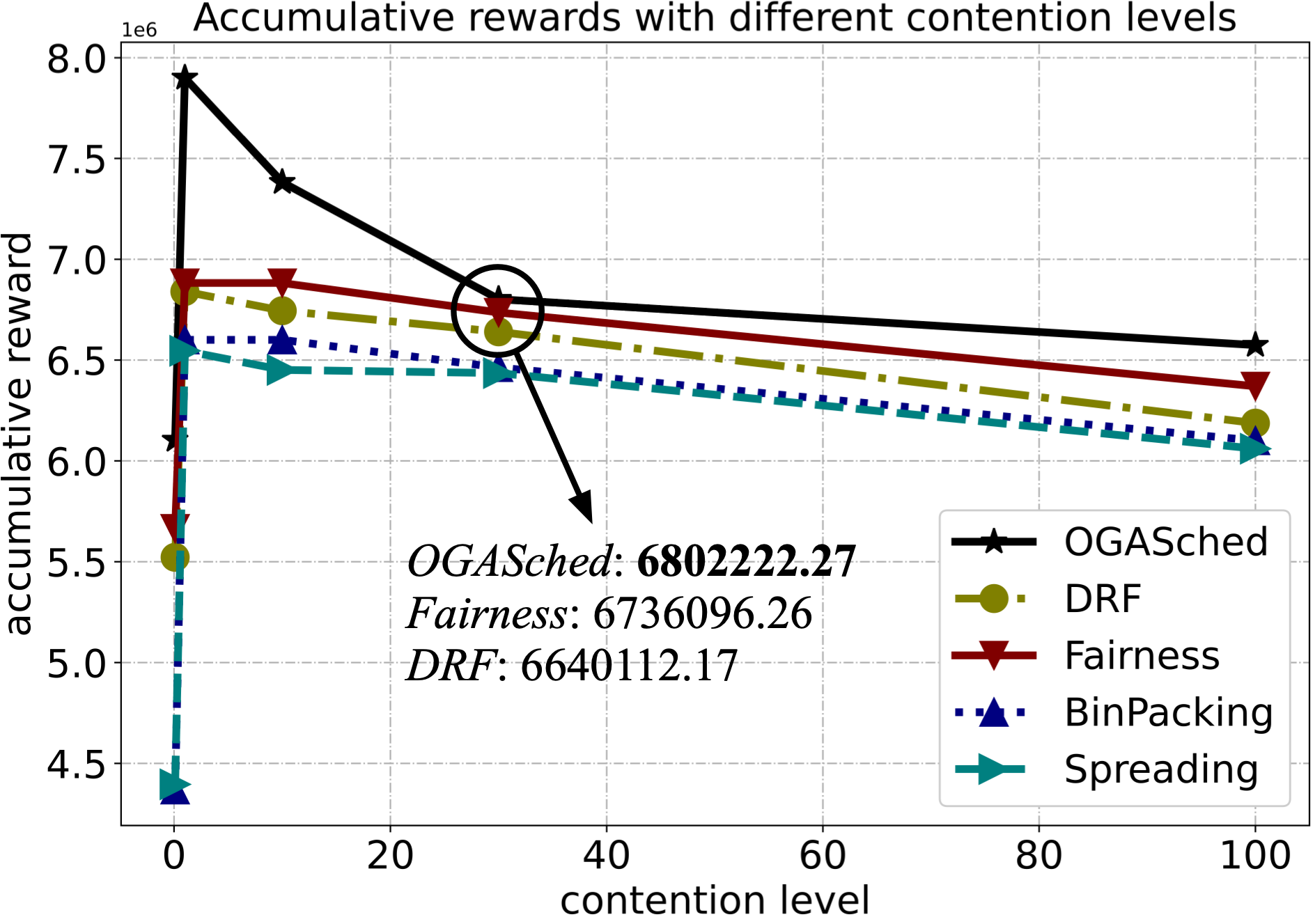}
    }
    \caption{Scalability verification of \textsc{OgaSched} under different scales of the bipartite graph and the contention levels.}
    \label{fig4}
\end{figure*}

\begin{table*}[h]
    \begin{center}
        \caption{\label{tab3}Generality and Robustness validation under different scenario settings.}   
        \begin{tabular}{c|cccc|cccc|ccc}
            \toprule
            & \multicolumn{4}{c|}{Time Horizon Length $T$} & \multicolumn{4}{c|}{Job Arrival Probability $\rho$} & \multicolumn{3}{c}{Graph Dense} \\
            \multirow{-2}{*}{Avg. Reward} & $1000$ & $2000$ & $5000$ & $10000$ & $0.3$ & $0.5$ & $0.7$ & $0.9$ & $\approx 2$ & $\approx 2.5$ & $\approx 3$\\ 
            \midrule
            
            \textsc{OgaSched} & 
                % T
                \textbf{2578.53} & 
                \textbf{2886.33} &
                \textbf{2911.37} & 
                \cellcolor[HTML]{EFEFEF}{\color[HTML]{000000}\textbf{3104.98}} & 
                % \rho
                \textbf{1904.87} & 
                \textbf{2154.18} & 
                \cellcolor[HTML]{EFEFEF}\textbf{3117.29} & 
                \textbf{2938.22} & 
                % Graph Dense
                \textbf{2816.18} & 
                \textbf{2904.51} & 
                \cellcolor[HTML]{EFEFEF}\textbf{3127.47} \\
            
            \textsc{DRF} & 
                % T
                2422.47 & 
                2493.02 & 
                2449.23 & 
                \cellcolor[HTML]{EFEFEF}\textbf{2497.85} &
                % \rho
                \textbf{1364.53} & 
               \textbf{2086.59} & 
                2503.01 & 
                \cellcolor[HTML]{EFEFEF} 2755.41 & 
                % Graph Dense
                2417.08 & 
                2786.94 & 
                \cellcolor[HTML]{EFEFEF}2795.42 \\
            
            \textsc{Fairness} & 
                % T
                \textbf{2532.24} & 
                \cellcolor[HTML]{EFEFEF}\textbf{2582.80} & 
                \textbf{2552.41} & 
                2436.22 & 
                % \rho
                1295.53 & 
                1997.19 & 
                \textbf{2628.02} & 
                \cellcolor[HTML]{EFEFEF}\textbf{2873.84} & 
                % Graph Dense
                \textbf{2501.54} & 
                \textbf{2857.60} & 
                \cellcolor[HTML]{EFEFEF}\textbf{2918.98} \\
            
            \textsc{BinPacking} & 
                % T
                2386.01 & 
                \cellcolor[HTML]{EFEFEF}2449.15 & 
                2444.32 & 
                2365.13 & 
                % \rho
                1246.39 & 
                \cellcolor[HTML]{EFEFEF}1897.79 & 
                2518.98 & 
                \cellcolor[HTML]{EFEFEF}2740.19 & 
                % Graph Dense
                2374.31 & 
                2757.71 & 
                \cellcolor[HTML]{EFEFEF}2829.19 \\
            
            \textsc{Spreading} & 
                % T
                2382.01 & 
                \cellcolor[HTML]{EFEFEF}2466.71 & 
                2436.60 & 
                2362.88 & 
                % \rho
                1250.67 & 
                1888.06 & 
                2519.37 & 
                \cellcolor[HTML]{EFEFEF}2737.93 & 
                % Graph Dense
                2382.87 & 
                2766.07 & 
                \cellcolor[HTML]{EFEFEF} 2836.37 \\ 
            
            \bottomrule
        \end{tabular}
    \end{center}
\end{table*}

In Fig. \ref{2-a}, the $y$-axis is the average reward unitl time $t$, i.e., $\frac{1}{t}\sum_{\tau=1}^t q \big(\vec{x}(\tau), \vec{y}(\tau) \big)$. Compared with the baselines, \textsc{OgaSched} has a clear advantage on the performance (with the increases of $11.33\%$, $7.75\%$, $13.89\%$, and $13.44\%$ compared with \textsc{DRF}, \textsc{Fariness}, \textsc{BinPacking}, and \textsc{Spreading}, respectively). Besides, it shows that the performance of \textsc{OgaSched} tends to increase as the length of the time horizon increases. The curve of \textsc{OgaSched} starts steep and later flattens. The reason is that, as a learning-powered algorithm, \textsc{OgaSched} learns the underlying distribution of job arrival patterns and it can make better decisions by adjusting the step directions. It is interesting to find that the rewards oscillate at the beginning time slots. One of the leading factors is that \textsc{OgaSched} is not boosted with a well-designed initial solution. In our experiments, a $8000$-time slot training only takes one hour. Thus, not surprisingly, the rewards achieved in the beginning can be easily surpassed when the time slot is sufficiently large.

It is not a surprise that \textsc{Fairness} achieves the best among the baselines. \textsc{Fairness} adopts a proportional allocation strategy and allocates resources to each non-empty port \textit{without bias}, which increases the computation gains adequately. When the contention is not fierce while the communication overhead is low, the advantages of \textsc{Fairness} will be more steady. By contrast, the advantages of \textsc{BinPacking} and \textsc{Spreading} are respectively high resource utilization and job isolation, which do not contribute to the reward directly. 

Fig. \ref{2-b} shows that the cumulative rewards achieved by all the five algorithms. In the beginning, \textsc{Fairness} and \textsc{DRF} have the slight edge, benefiting by the propotional allocation idea. Nevertheless, as the time slot increases, \textsc{OgaSched} is able to surpass them without difficulty. Fig. \ref{2-c} demonstrates the ratio on the achieved average rewards between \textsc{OgaSched} and the baselines. Similarly, the ratios oscillate at the beginning. After that, they increase steeply and later flattens. 

\begin{figure}[h]
    \centering 
    \subfigure[Impact of $\eta_0$.]{ 
        \label{3-a}
        \includegraphics[height=1.1in]{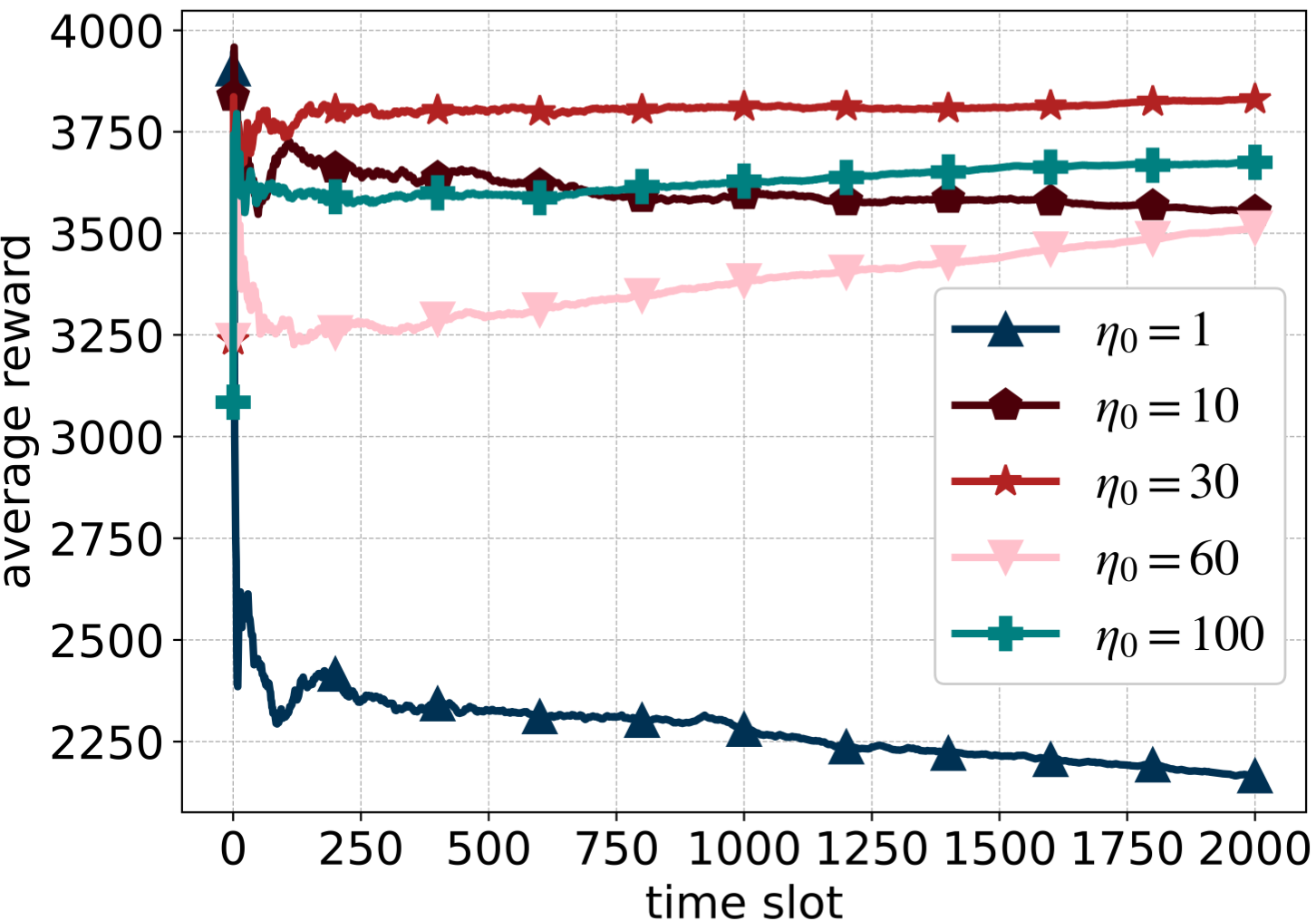}
    }
    \subfigure[Impact of decay.]{ 
        \label{3-b}
        \includegraphics[height=1.1in]{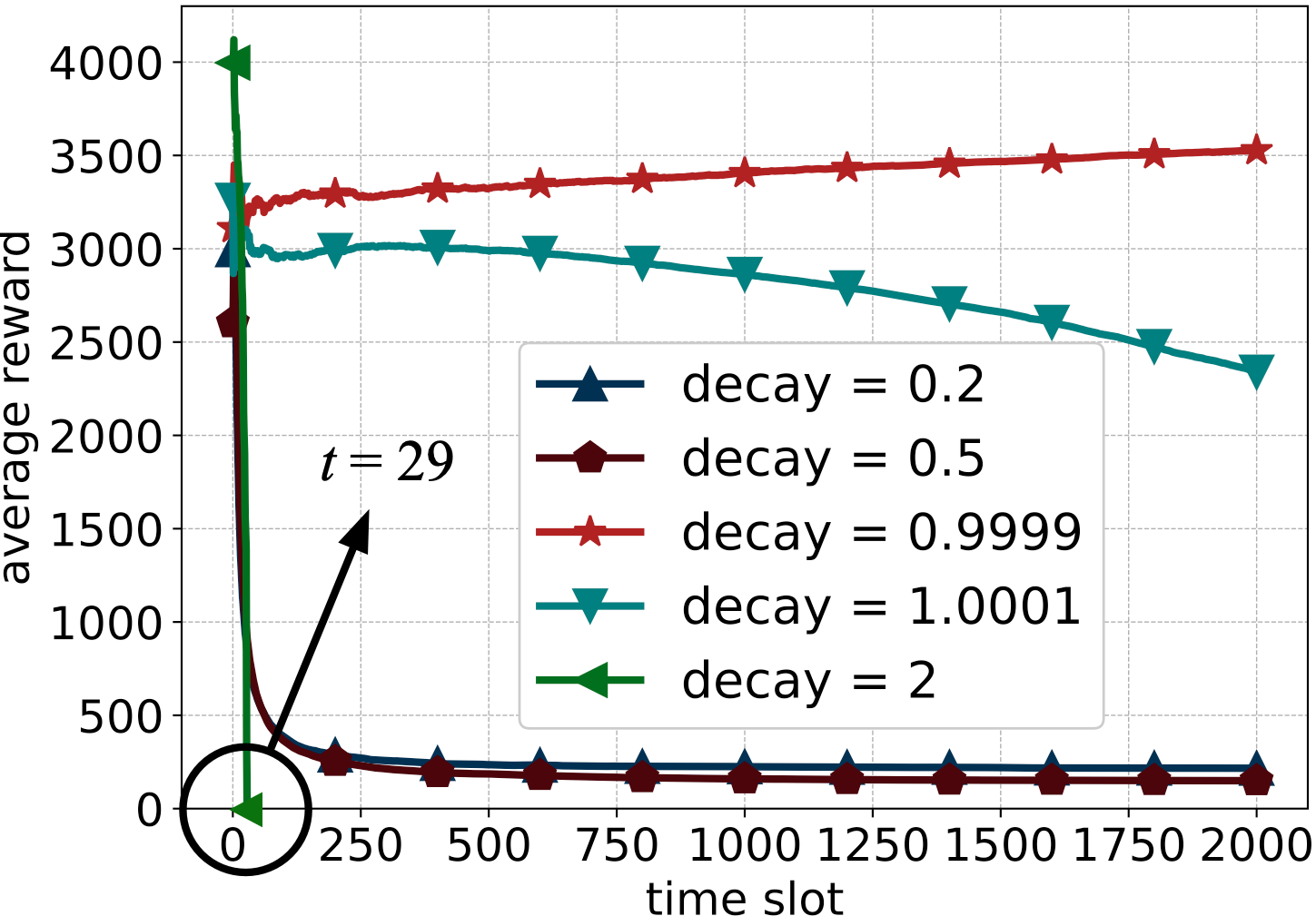}
    }
    \caption{The performance of \textsc{OgaSched} with different hyper-parameters.}
    \label{fig3}
\end{figure}

The hyper-parameters of \textsc{OgaSched}, especially the initial learning rate $\eta_0$ and the decay, have a remarkable impact on its performance. From Fig. \ref{fig3} we can find that, a wrong setting of these hyper-parameters could lead to a poor performance, even the decrease of the cumulative reward (which means, the reward is negative in many time slots). At the last of Sec. \ref{s3.3}, we claim that, to achieve an cumulative reward with a lower bound guarantee, at each time $t$, the learning rate should be set around $\eqref{set-eta}$. Note that in $\eqref{set-eta}$, the learning rate is encouraged to be larger and larger as time moves, which is counterintuitive and it goes against the convergence to a local optimum. The curves in Fig. \ref{3-b} also verify that, setting decay as $0.9999$ is better than $1.0001$. The best decay in practice does not follow the guidance of theory because the regret analysis only gives the \textit{worst} case guarantee on the cumulative rewards. In our experiments, the best range for decay is $[0.995, 0.9999]$. 

\subsection{Scalability, Generality and Robustness Evaluations}\label{s4.2}

\begin{figure*} 
    \centering 
    \subfigure[Average rewards.]{ 
        \label{5-a}
        \includegraphics[height=1.48in]{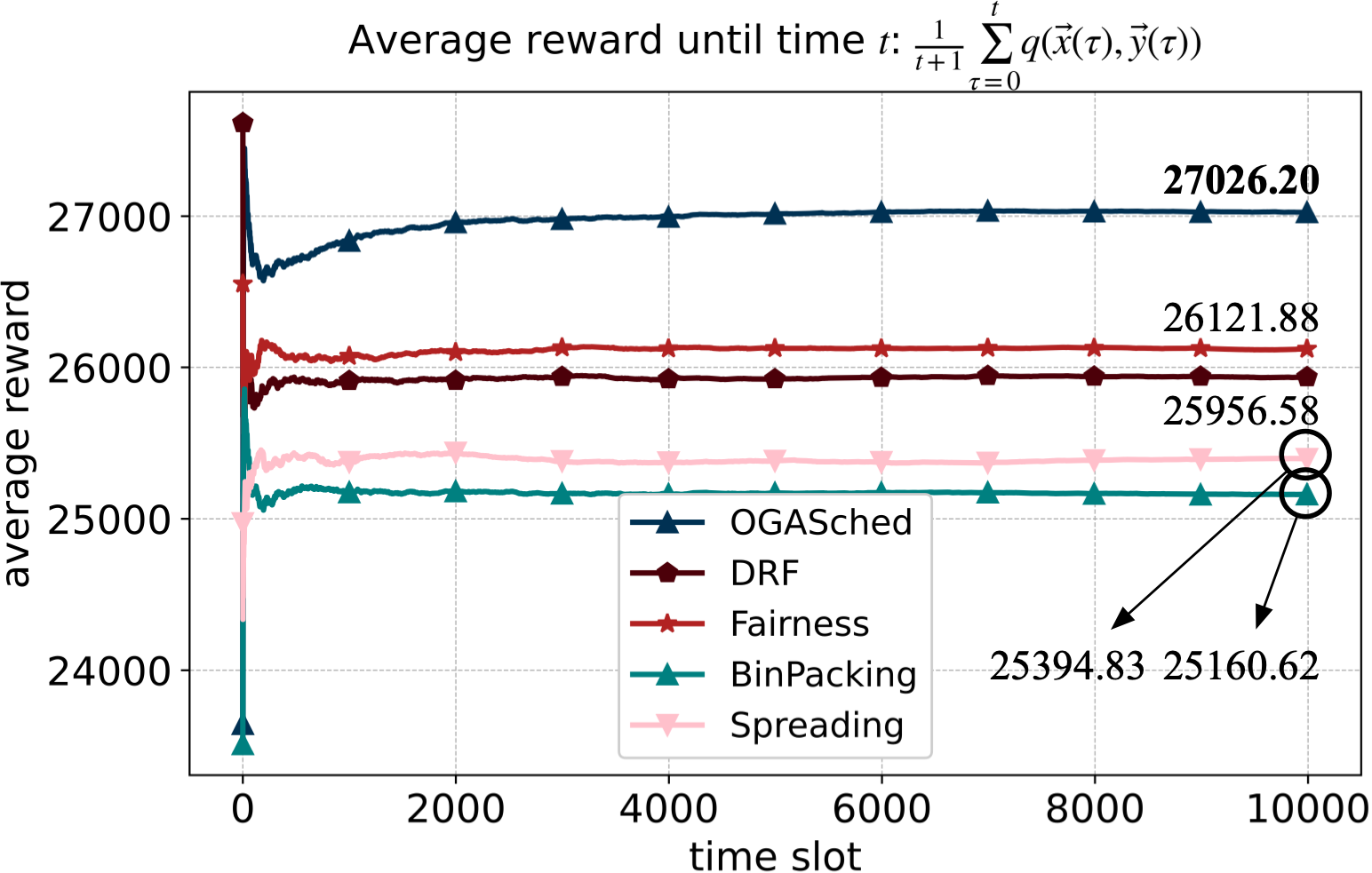}
    }
    \subfigure[Accumulative rewards.]{ 
        \label{5-b}
        \includegraphics[height=1.48in]{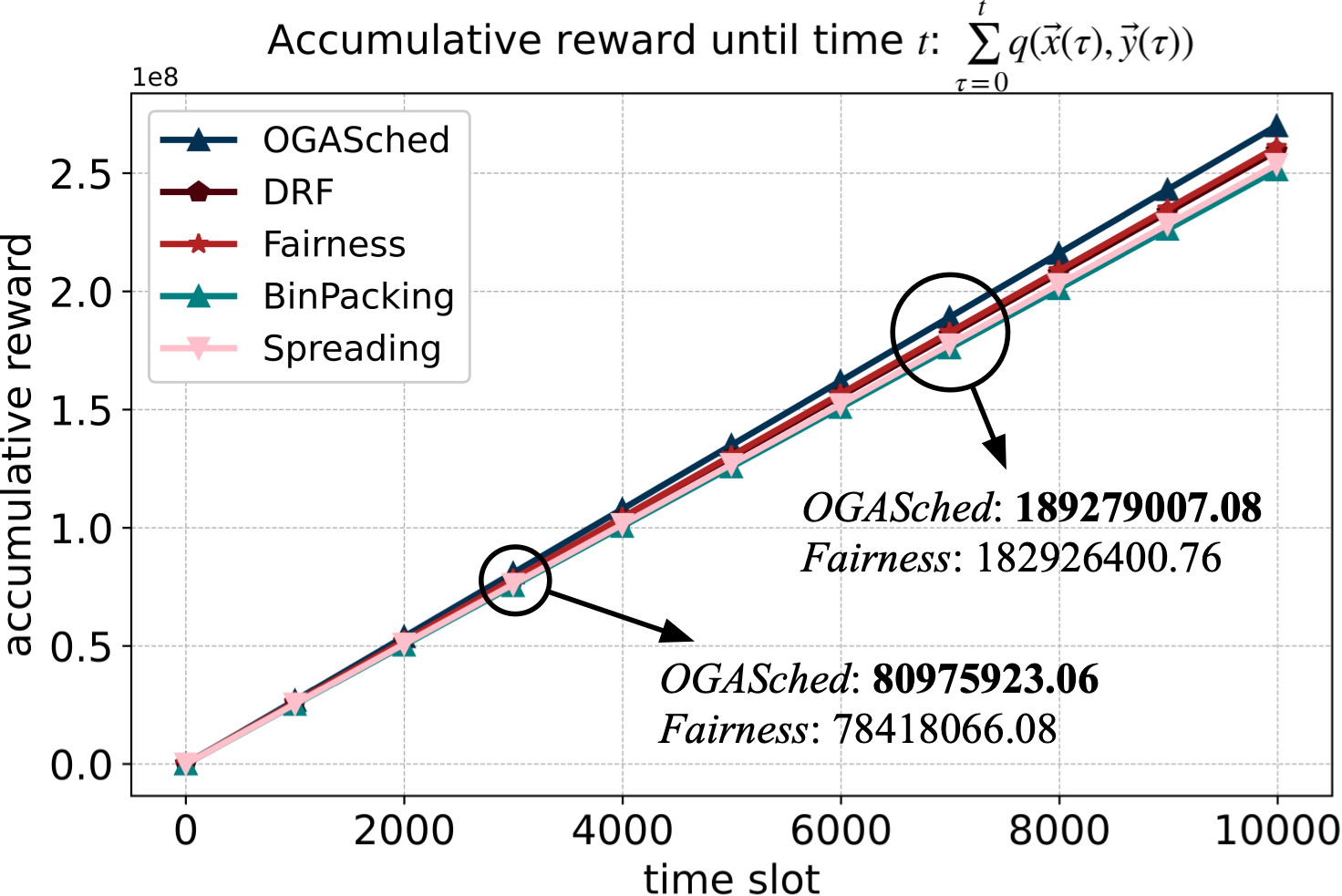}
    }
    \subfigure[Ratios over the baselines.]{ 
        \label{5-c}
        \includegraphics[height=1.48in]{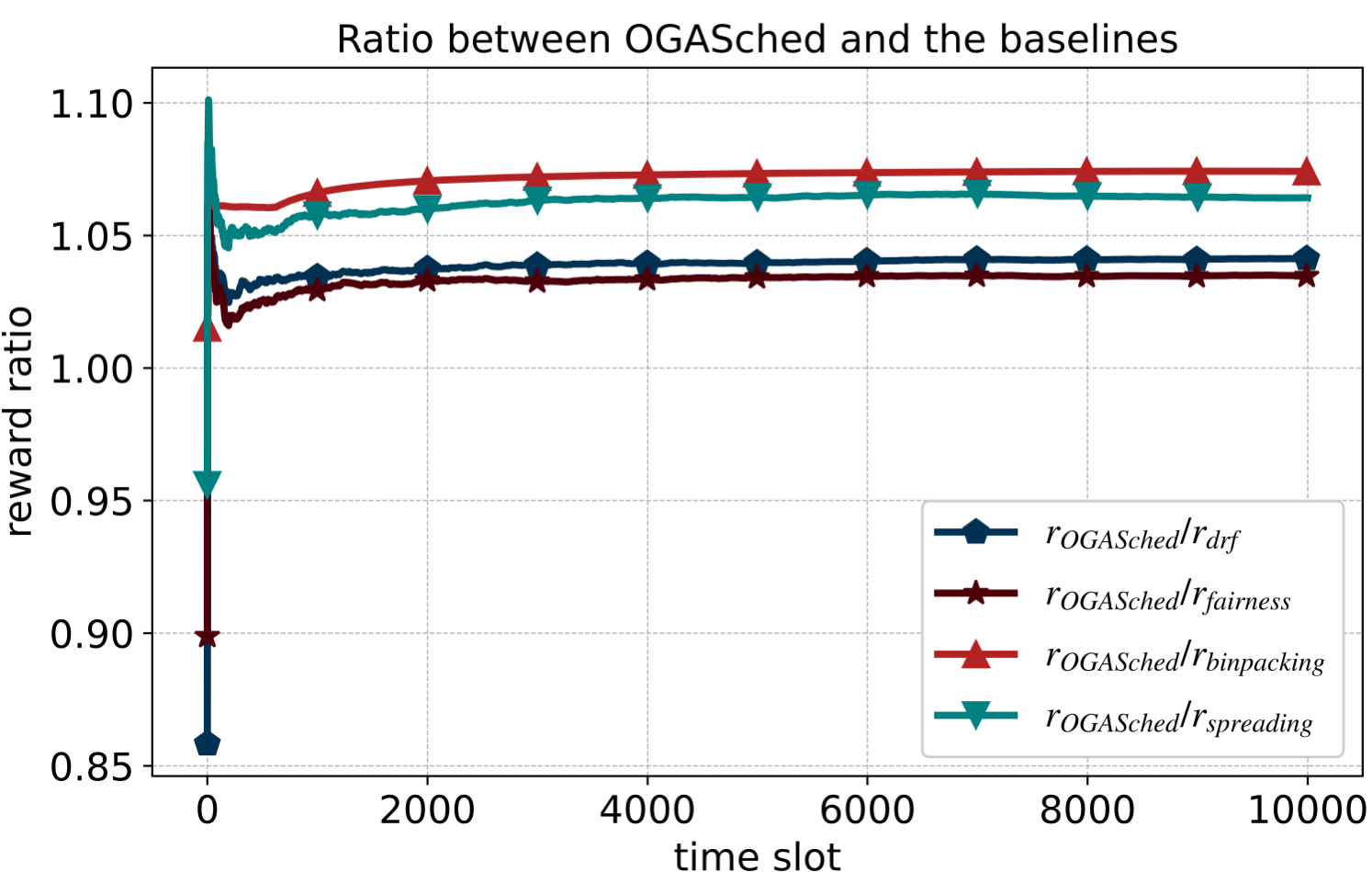}
    }
    \caption{Large-scale validations. It takes 15 hours for \textsc{OgaSched} to complete when $T = 10000$, $\beta \in [0.01, 0.015]$, and contention level is $5$.}
    \label{fig5}
\end{figure*}

\begin{figure}[htbp]
    \centerline{\includegraphics[width=2.8in]{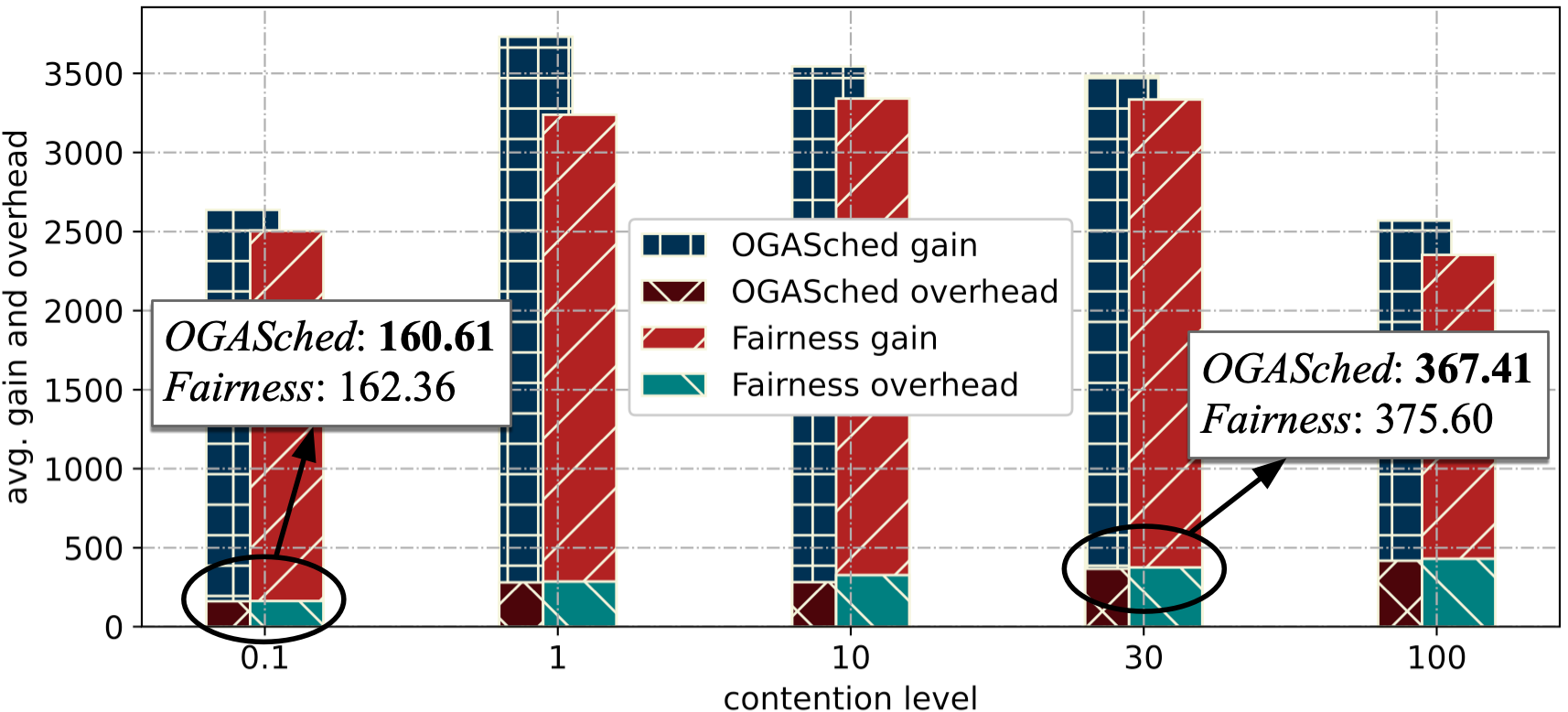}}
    \caption{Average computation gain and communication overhead of each time slot under different contention levels.}
    \label{fig-gain-overhead}
\end{figure}

\begin{figure}[htbp]
    \centerline{\includegraphics[width=2.8in]{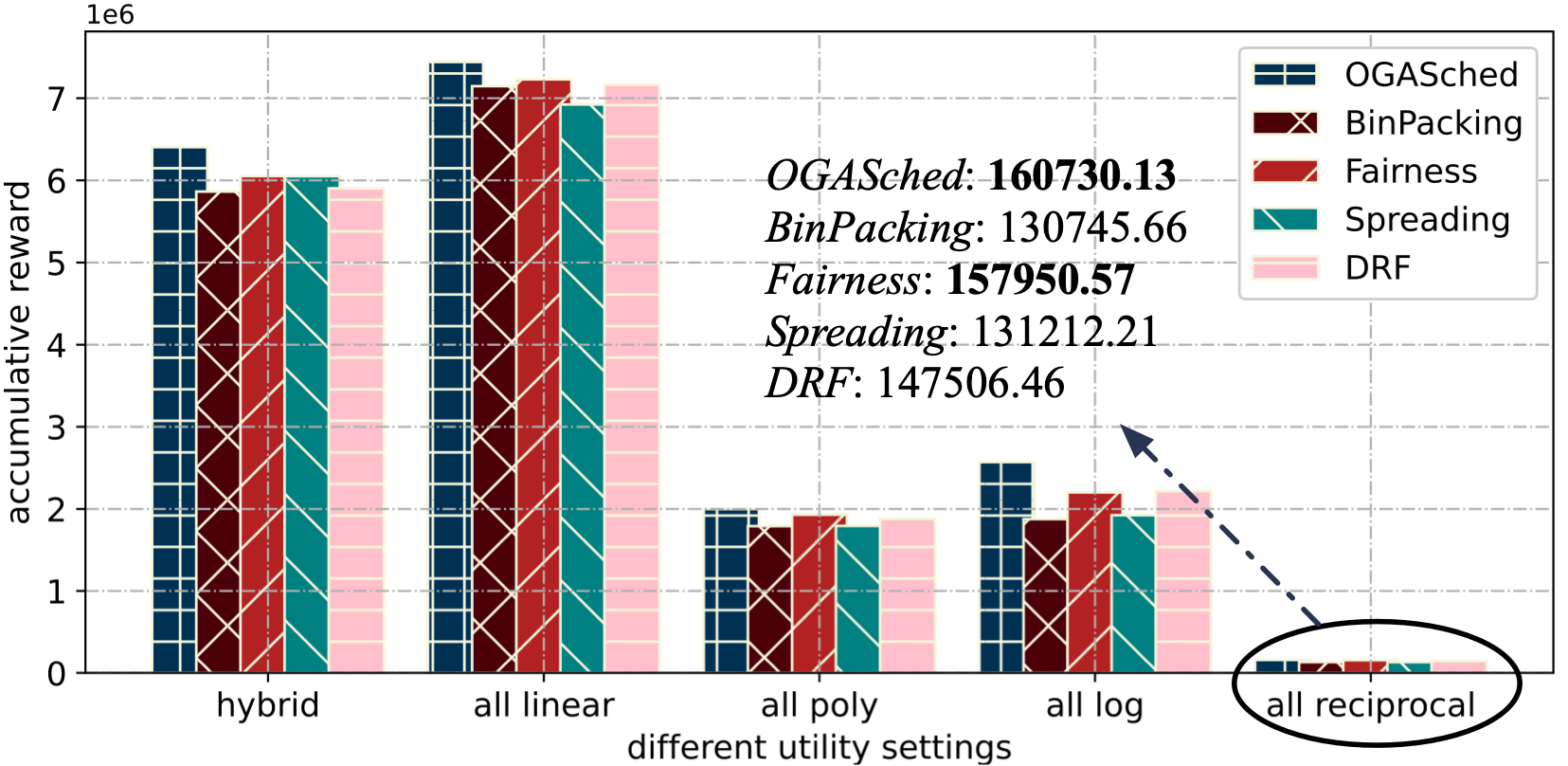}}
    \caption{Accumulative rewards with different utilities.}
    \label{fig-utility}
\end{figure}
In this section, we evaluate the performance of \textsc{OgaSched} under different scales of scenario settings. Fig. \ref{4-a} and Fig. \ref{4-b} demonstrate the impact of the scale of the bipartite graph $\mathcal{G}$. In these two figures, the left $y$-axis is the cumulative reward while the right $y$-axis is the ratio $r_a / r_b$, where $r_a$ is the cumulative reward achieved by \textsc{OgaSched}, and $r_b$ is the baselines'. Firstly, we observe that, whatever the number of the computing instances is, \textsc{OgaSched} takes the leading position. Besides, as $|\mathcal{R}|$ increases, all the algorithms obtain a larger cumulative reward. The result is evident because a large cluster can provide sufficient resources, which leads to jobs being fully served. It is also worth noting that, when $|\mathcal{R}|$ increases, the superiority of \textsc{OgaSched} over the baselines firstly increases then decreases. It demonstrates that the resource contention is fierce when $|\mathcal{R}| \in [128, 256]$. In this case, it is necessary for \textsc{OgaSched} to be trained with a larger time slot. Fig. \ref{4-b} shows that the number of job types, i.e., $|\mathcal{L}|$, has a weaker impact than $|\mathcal{R}|$ to the performance of \textsc{OgaSched}. The phenomenon verifies the conclusion we have concluded, i.e., the regret grows linearly with $|\mathcal{R}|$, but it is sublinear with $|\mathcal{L}|$. 

Fig. \ref{4-c} shows the impact of contention level. This parameter works as a multiplier to the resource requirements of jobs. We can observe that, when moving contention level from $0.1$ to $1$, all the achieved cumulative rewards increase. This is obvious because  a larger resource requirement leads to a larger computation gain on the premise of low contention. However, increasing the multiplier further leads to the downgrade of performances and the reduction of the superiority of \textsc{OgaSched}. Even so, \textsc{OgaSched} always performs the best. Fig. \ref{fig-gain-overhead} shows the average computation gain and communication overhead penalty of each time slot under different contention levels. We can find that the penalty increases with the contention level slowly.
 
Fig. \ref{fig-utility} demonstrates the cumulative rewards with different utilities. Because of the diminishing marginal effect, the rewards with \textit{ploy}, \textit{log}, and \textit{reciprocal} utilities are significantly less than the rewards with \textit{linear} utilities. Nevertheless, the diminishing marginal effect does not change the superiority of \textsc{OgaSched} against the baselines. Even in the \textit{all reciprocal} utility settings, for \textsc{Fairness}, \textsc{OgaSched} has its advantages. 

In addition to the above evaluations, we also test the generality and robustness of \textsc{OgaSched} under different settings of the following parameters: the time horizon length $T$, the job arrival probability $\rho$, and the dense of the bipartite graph. The graph dense is calculated as $\sum_{r \in \mathcal{R}} |\mathcal{L}_r| / |\mathcal{R}|$. The results are shown in Tab. \ref{tab3}. The two largest values in each column of the table are made bold. Besides, for each parameter and each algorithm, the setting which leads to the largest reward is marked with a light-grey background. We summarize the key findings as follows.

\begin{itemize}
    \item Firstly, whatever the parameter settings, \textsc{OgaSched} always performs the best, and its performance has a positive correlation with the 
    time horizon length $T$. As we have analyzed, a large time horizon provides more chances for \textsc{OgaSched} to learn the underlying distributions, 
    thereby increasing the reward in the gradient ascent directions.
    \item Increasing the job arrival probability can lead to a high resource utilization, thereby increasing the rewards. However, a large job arrival 
    probability also brings in a fierce resource contention. A direct consequence of it is that, for \textsc{OgaSched}, many elements in the 
    vector $\vec{y}(t)$ fall into the interior of $\mathcal{Y}$, rather than the boundaries, thereby leading to a reward reduction. The phenomenon 
    can be observed when moving $\rho$ from $0.7$ to $0.9$.
    \item Graph dense has a similar effect on the reward to the job arrival probability. Nevertheless, the reasons behind are distinct. 
    A larger graph dense increases the opportunities for a job to be served with a large possible parallelism, thereby increasing the 
    computation gain. By contrast, the communication overhead has a slow rate of growth. 
\end{itemize}

\subsection{Large-Scale Validations}\label{s4.3}
To test the efficacy of \textsc{OgaSched} in large-scale scenarios, we conduct the following experiments. In these experiments, the number of the 
job types is set as $100$ while the quantity of the computing instances is $1024$ in default. The results in Fig. \ref{fig5} show that the superiority 
of \textsc{OgaSched} is preserved even in large-scale scenarios. 
% ======================================================================================================================================================

\section{Related Works}\label{s5}
The design of online job scheduling algorithms that yield a nice theoretical bound is always the focus of attention from the research community. 
Existing online job scheduling algorithms can be organized into two categories.

In the first category, the online algorithms are elaborately designed for specific job types, such as DNN model training \cite{pollux,8486422,9328612,bao2018online,yu2021sum,BSP,narayanan2020heterogeneity}, big-data query \& analytics \cite{decima,map-reduce}, multi-stage workflows \cite{8416357,8486340,8710674,8746711},
etc. A typical work on DNN model training is \cite{yu2021sum}, where the authors fully take the layered structure of DNNs into consideration and 
develop an efficient resource scheduling algorithm based on the sum-of-ratios multi-type-knapsack decomposition method. The authors further 
prove that the proposed algorithm has a SOTA approximation ratio within a polynomial running time. \cite{BSP} is another work that fully explores the 
Bulk Synchronous Parallel (BSP) property of the DNN training jobs. The authors develop an algorithm which is $\mathcal{O} (\ln |\mathcal{M}|)$-approximate 
with high probability, where $\mathcal{M}$ is the set of resources. These works are designed for specific job types, and they do not 
provide a general analysis of the gain-overhead tradeoff for multi-server jobs. This paper intends to fill the gap. 

In the second category, the types of job are not specified, while the theoretical superiority is highlighted. The algorithms are designed with 
different theoretical basis, including online approximate algorithms \cite{8941266,8737612,8917749}, Online Convex Optimization (OCO) techniques 
\cite{8737465}, game-theoretical approaches \cite{8737370}, online learning and DRL-based algorithms \cite{liang2020data,9488701}, etc. 
In these works, the performance of the proposed algorithms is usually analyzed with approximate ratio, competitive ratio, Price of Anarchy (PoA), 
and regret. A typical recent work is \cite{8737465}. The authors develop an algorithm whose dynamic regret is upper bounded by 
$\mathcal{O} (\textsc{Opt}^{1-\beta})$, where $\beta \in [0, 1)$. None of the existing works analyze the gain-overhead tradeoff and provide a regret 
of $\mathcal{O}(\sqrt{|\mathcal{L}| T})$ as this paper demonstrates. 
% ======================================================================================================================================================

\section{Conclusions}\label{s6}
In this paper, we study the online scheduling of multi-server jobs in terms of the gain-overhead tradeoff. The problem is 
formulated as an cumulative reward maximization program. The reward of scheduling a job is designed as the difference between the 
computation gain and the penalty on the dominant communication overhead. We propose an algorithm, i.e. \textsc{OgaSched}, to 
learn the best possible scheduling decision in the ascending direction of the reward gradients. \textsc{OgaSched} is the first 
algorithm that has a sublinear regret w.r.t. the number of job types and time slot length, which is a SOTA result for 
concave rewards. \textsc{OgaSched} is well designed to be parallelized, which makes large-scale applications possible. The superiority 
of \textsc{OgaSched} is also validated with extensive trace-driven simulations. Future extensions may include, i.e., more 
elaborate modeling and analysis of the intra-node and inter-node communication overheads.

% ======================================================================================================================================================

\ifCLASSOPTIONcompsoc
  \section*{Acknowledgments}
\else
  \section*{Acknowledgment}
\fi
This work was supported in part by the National Key Research and Development Program of China under Grant 2022YFB4500100, the  National Science Foundation of China under Grants 62125206 and U20A20173, and the Key Research Project of Zhejiang Province under Grant 2022C01145.

\bibliographystyle{IEEEtran}
\bibliography{IEEEabrv,ref.bib}

\begin{IEEEbiography}
    [{\includegraphics[width=1in,height=1.25in,clip,keepaspectratio]{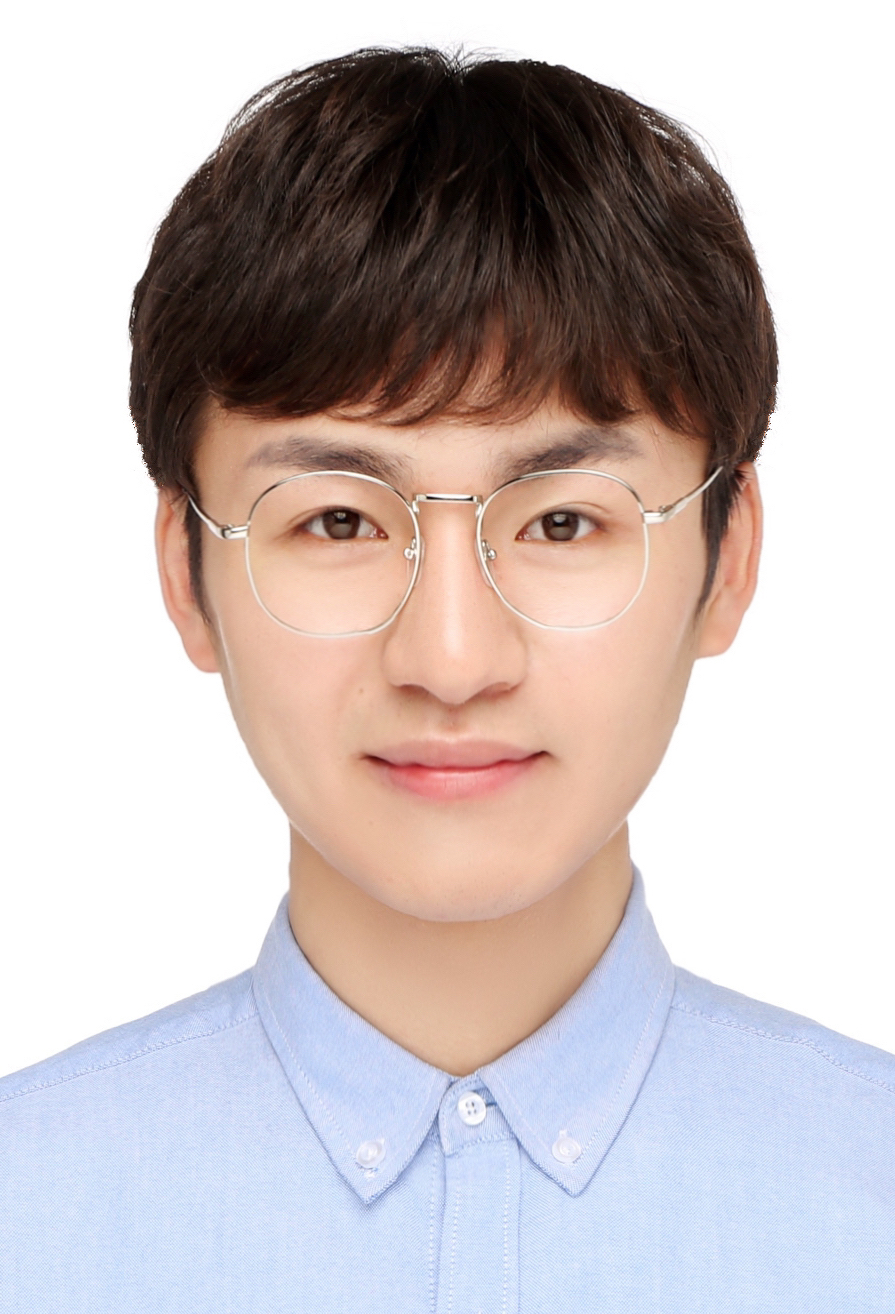}}]{Hailiang Zhao} received the B.S. degree in 
    2019 from the school of computer science and technology, Wuhan University of Technology, Wuhan, China. He is currently pursuing the 
    Ph.D. degree with the College of Computer Science and Technology, Zhejiang University, Hangzhou, China. His research interests include 
    cloud \& edge computing, distributed systems and optimization algorithms. He has published several papers in flagship conferences 
    and journals including IEEE ICWS 2019, IEEE TPDS, IEEE TMC, etc. He has been a recipient of the Best Student Paper Award of IEEE ICWS 2019. 
    He is a reviewer for IEEE TSC and Internet of Things Journal.
\end{IEEEbiography}

\begin{IEEEbiography}
    [{\includegraphics[width=1in,height=1.25in,clip,keepaspectratio]{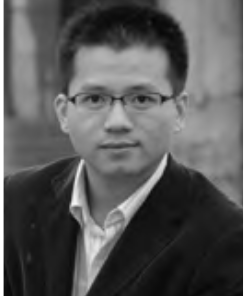}}]{Shuiguang Deng} 
    is currently a full professor at the College of Computer Science and Technology in Zhejiang University, China, 
    where he received a BS and PhD degree both in Computer Science in 2002 and 2007, respectively. He previously 
    worked at the Massachusetts Institute of Technology in 2014 and Stanford University in 2015 as a visiting scholar. 
    His research interests include Edge Computing, Service Computing, Cloud Computing, and Business Process Management. 
    He serves for the journal IEEE Trans. on Services Computing, Knowledge and Information Systems, Computing, and IET 
    Cyber-Physical Systems: Theory \& Applications as an Associate Editor. Up to now, he has published more than 100 
    papers in journals and refereed conferences. In 2018, he was granted the Rising Star Award by IEEE TCSVC. He is 
    a fellow of IET and a senior member of IEEE.
\end{IEEEbiography}

\begin{IEEEbiography}
    [{\includegraphics[width=1in,height=1.25in,clip,keepaspectratio]{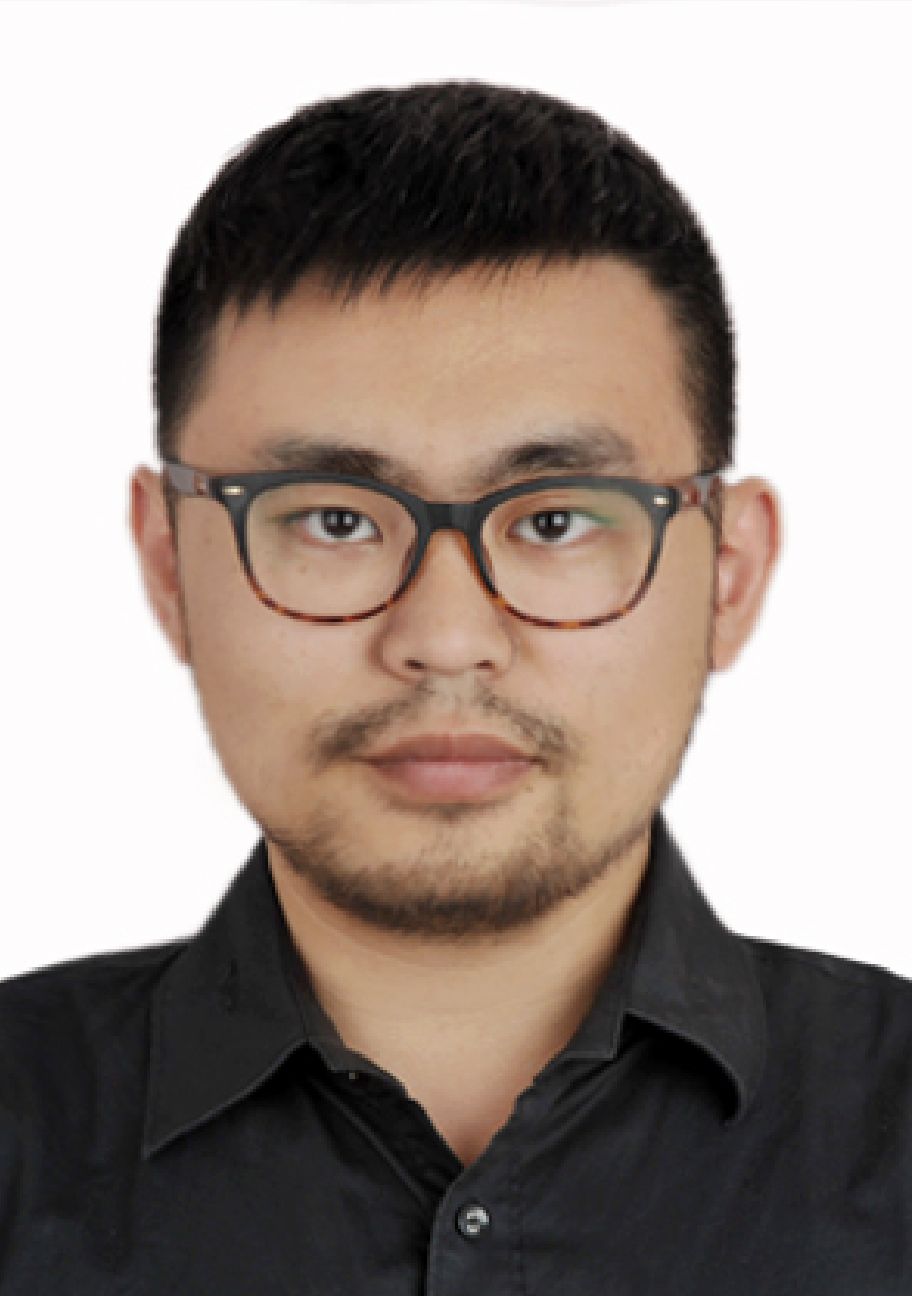}}]{Zhengzhe Xiang}
    received the B.S. and Ph.D. degree of Computer Science and Technology in Zhejiang University, Hangzhou, 
    China. He was previously a visiting student worked at the Karlstad University, Sweden in 2018. He is 
    currently a Lecturer with Zhejiang University City College, Hangzhou, China. His research interests lie in 
    the fields of Service Computing, Cloud Computing, and Edge Computing. 
\end{IEEEbiography}

\begin{IEEEbiography}
    [{\includegraphics[width=1in,height=1.25in,clip,keepaspectratio]{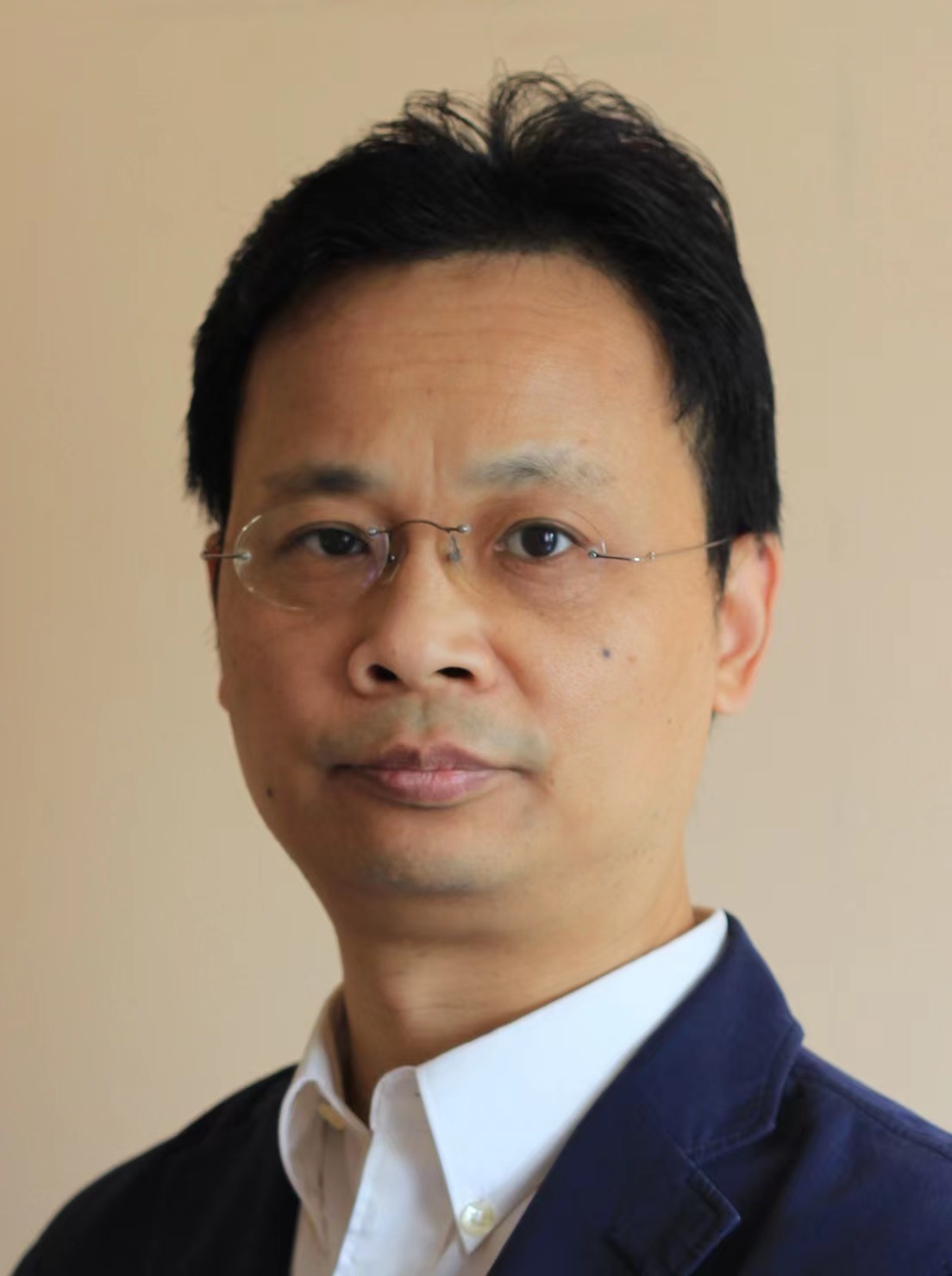}}]{Xueqiang Yan}
    is currently a technology expert with the Wireless Technology Lab, Huawei Technologies. He was a member of technical staff at Bell Labs from 2000 to 2004. From 2004 to 2016, he was the director of the Strategy Department, Alcatel-Lucent Shanghai Bell. His current research interests include wireless networking, the Internet of Things, edge AI, future mobile network architecture, network convergence, and evolution.
\end{IEEEbiography}

\begin{IEEEbiography}
    [{\includegraphics[width=1in,height=1.25in,clip,keepaspectratio]{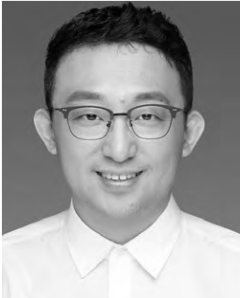}}]{Jianwei Yin} 
    received the Ph.D. degree in computer science from Zhejiang University (ZJU) in 2001. 
    He was a Visiting Scholar with the Georgia Institute of Technology. He is currently a Full Professor 
    with the College of Computer Science, ZJU. Up to now, he has published more than 100 papers in top 
    international journals and conferences. His current research interests include service computing 
    and business process management. He is an Associate Editor of the IEEE Transactions on Services 
    Computing.
\end{IEEEbiography}

\begin{IEEEbiography}[{\includegraphics[width=1in,height=1.25in,clip,keepaspectratio]{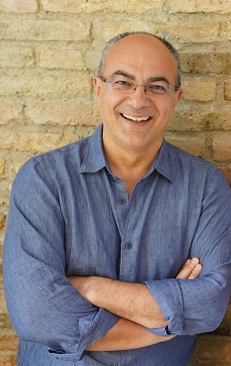}}]{Schahram Dustdar}
    is a Full Professor of Computer Science (Informatics) with a focus on Internet Technologies heading the Distributed 
    Systems Group at the TU Wien. He is founding co-Editor-in-Chief of ACM Transactions on Internet of Things (ACM TIoT) as well as Editor-in-Chief of Computing (Springer). He is an Associate Editor of IEEE Transactions on Services Computing, IEEE Transactions on Cloud Computing, ACM Computing Surveys, ACM Transactions on the Web, and ACM Transactions on Internet Technology, as well as on the editorial board of IEEE Internet Computing and IEEE Computer. Dustdar is recipient of multiple awards: TCI Distinguished Service Award (2021), IEEE TCSVC Outstanding Leadership Award (2018), IEEE TCSC Award for Excellence in Scalable Computing (2019), ACM Distinguished Scientist (2009), ACM Distinguished Speaker (2021), IBM Faculty Award (2012). He is an elected member of the Academia Europaea: The Academy of Europe, where he is chairman of the Informatics Section, as well as an IEEE Fellow (2016), an Asia-Pacific Artificial Intelligence Association (AAIA) President (2021) and Fellow (2021). He is an EAI Fellow (2021) and an I2CICC Fellow (2021). He is a Member of the 2022 IEEE Computer Society Fellow Evaluating Committee (2022).

\end{IEEEbiography}

\begin{IEEEbiography}[{\includegraphics[width=1in,height=1.25in,clip,keepaspectratio]{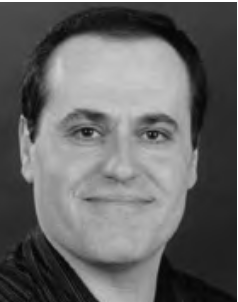}}]{Albert Y. Zomaya}
    is the Peter Nicol Russell Chair Professor of Computer Science and Director of the Centre for Distributed 
    and High-Performance Computing at the University of Sydney. To date, he has published > 600 scientific papers and articles and is (co-)author/editor 
    of > 30 books. A sought-after speaker, he has delivered > 250 keynote addresses, invited seminars, and media briefings. His research interests 
    span several areas in parallel and distributed computing and complex systems. He is currently the Editor in Chief of the ACM Computing Surveys 
    and processed in the past as Editor in Chief of the IEEE Transactions on Computers (2010-2014) and the IEEE Transactions on Sustainable Computing (2016-2020).
    
    Professor Zomaya is a decorated scholar with numerous accolades including Fellowship of the IEEE, the American Association for the Advancement 
    of Science, and the Institution of Engineering and Technology (UK). Also, he is an Elected Fellow of the Royal Society of New South Wales and 
    an Elected Foreign Member of Academia Europaea. He is the recipient of the 1997 Edgeworth David Medal from the Royal Society of New South Wales 
    for outstanding contributions to Australian Science, the IEEE Technical Committee on Parallel Processing Outstanding Service Award (2011), 
    IEEE Technical Committee on Scalable Computing Medal for Excellence in Scalable Computing (2011), IEEE Computer Society Technical Achievement 
    Award (2014), ACM MSWIM Reginald A. Fessenden Award (2017), the New South Wales Premier’s Prize of Excellence in Engineering and Information 
    and Communications Technology (2019), and the Research Innovation Award, IEEE Technical Committee on Cloud Computing (2021). 
  \end{IEEEbiography}

\end{document}